\documentclass[fleqn,usenatbib]{stile/mnras}

\usepackage{txfonts}
\usepackage[T1]{fontenc}
\usepackage{ae,aecompl}

\usepackage{graphicx}

\usepackage{amsmath}
\usepackage{amsfonts}
\usepackage{amssymb}

\usepackage{xcolor}

\usepackage{breakcites}
\usepackage{hyperref}
\usepackage{footnote}

\usepackage{comment}

\usepackage{xfrac}        

\def\nat{Nature}

\def\mnras{MNRAS}
\def\apj{ApJ}
\def\apjl{ApJL}
\def\apjs{ApJS}
\def\aap{A\&A}

\def\aapr{A\&A Rev.}
\def\araa{ARA\&A}

\def\pasj{Pub. Astron. Soc. Japan}

\newcommand{\quotes}[1]{``#1''}
\newcommand\code[1]{\textsc{\MakeLowercase{#1}}}

\newcommand{\angstrom}{\textup{\AA}}

\def\gtsima{$\; \buildrel > \over \sim \;$}
\def\ltsima{$\; \buildrel < \over \sim \;$}
\def\gsim{\lower.5ex\hbox{\gtsima}}
\def\lsim{\lower.5ex\hbox{\ltsima}}
\def\gtsima{$\; \buildrel > \over \sim \;$} 
\def\ltsima{$\; \buildrel < \over \sim \;$} \def\gsim{\lower.5ex\hbox{\gtsima}} 
\def\lsim{\lower.5ex\hbox{\ltsima}} 
\def\simgt{\lower.5ex\hbox{\gtsima}} 
\def\simlt{\lower.5ex\hbox{\ltsima}}

\def\be{\begin{equation}}
\def\ee{\end{equation}}

\def\NII{\hbox{[N~$\scriptstyle\rm II $]}}

\def\Hb{\hbox{H$\beta $~}}
\def\Ha{\hbox{H$\alpha $~}}

\def\OII{\hbox{[O~$\scriptstyle\rm II $]}}
\def\CII{\hbox{[C~$\scriptstyle\rm II $]}}

\def\NIII{\hbox{[N~$\scriptstyle\rm III $]}}
\def\OIII{\hbox{[O~$\scriptstyle\rm III $]}}

\graphicspath{{plots/}}                 

\begin{document}

\title[Infrared properties of $z\sim 5$ galaxies]{A new look at the infrared properties of $z\sim 5$ galaxies}

\author[Sommovigo et al.]{L. Sommovigo$^{1}$\thanks{\href{mailto:laura.sommovigo@sns.it}{laura.sommovigo@sns.it}}, A. Ferrara$^{1}$, S. Carniani$^{1}$, A. Pallottini$^{1}$, P. Dayal$^{2}$,\newauthor E. Pizzati$^{3}$, M. Ginolfi$^{4}$,  V. Markov$^{1}$, A. Faisst$^{4}$\\
$^{1}$ Scuola Normale Superiore, Piazza dei Cavalieri 7, I-56126 Pisa, Italy\\
$^{2}${Kapteyn Astronomical Institute, University of Groningen, 9700 AV Groningen, The Netherlands}\\
$^{3}$ Leiden Observatory, Leiden University, PO Box 9500, 2300 RA Leiden, The Netherlands\\
$^{4}$ European Southern Observatory, Karl-Schwarzschild-Strasse 2, 85748 Garching, Germany\\
$^{5}$IPAC, M/C 314-6, California Institute of Technology, 1200 East California Boulevard, Pasadena, CA 91125, USA\\
}

\label{firstpage}
\pagerange{\pageref{firstpage}--\pageref{lastpage}}

\maketitle

\begin{abstract}
Recent ALMA large surveys unveiled the presence of significant dust continuum emission in star-forming galaxies at $z>4$. Unfortunately, such large programs  -- i.e. ALPINE ($z\sim 5$) and REBELS ($z \sim 7$) -- only provide us with a single Far-Infrared (FIR) continuum data point for their individual targets. 
Therefore, high-$z$ galaxies FIR spectral energy densities (SEDs) remain mostly unconstrained, hinging on an assumption for their dust temperature ($T_{\rm d}$) in the SED fitting procedure. This introduces uncertainties in the inferred dust masses ($M_{\rm d }$), infrared luminosities ($L_{\rm IR}$), and obscured Star Formation Rate (SFR) fraction at $z > 4$.
In this work we use a method that allows us to constrain $T_{\rm d}$ with a single band measurement by combining the $158\ \mathrm{\mu m}$ continuum information with the overlying \CII\ emission line. We analyse the $21$ \CII\ and FIR continuum detected $z\sim 5$ galaxies in ALPINE, finding a range of $T_{\rm d}=25-60\ \mathrm{K}$ and $M_{\rm d} = 0.6-25.1\ \times 10^{7}\ \mathrm{M_{\odot}}$. 
Given the measured stellar masses of ALPINE galaxies, the inferred dust yields are around $M_{\rm d}/M_{\star} = (0.2-8) \times 10^{-3}$, consistent with theoretical dust-production constraints.
We find that $8$ out of $21$ ALPINE galaxies have $L_{\rm IR} \geq 10^{12}\ \mathrm{L_{\odot}}$, comparable to UltraLuminous IR Galaxies (ULIRGs). Relying on ultraviolet-to-optical SED fitting, the SFR was underestimated by up to $2$ orders of magnitude in $4$ of these $8$ ULIRGs-like galaxies. We conclude that these $4$ peculiar sources should be characterised by a two-phase interstellar medium structure with \quotes{spatially-segregated} FIR and ultraviolet emitting regions. 

\end{abstract}


\section{Introduction}
The Atacama Large Millimeter Array (ALMA) opened a new window on the rest-frame far-infrared (FIR) emission of the first generations of galaxies, dramatically improving our understanding of the dust build-up in the early Universe. Dust grains shape the galaxies Spectral Energy Distributions (SEDs), by absorbing the stellar ultraviolet (UV) and optical radiation, and thermally re-emitting at mid-infrared (MIR, rest-frame $5-50\ \mathrm{\mu m}$) and far-infrared (rest-frame $50-1000\ \mathrm{\mu m}$) wavelengths \citep{draine1989interstellar,meurer1999dust,calzetti2000dust,weingartner2001dust,Draine03}. 
Dust FIR emission 
is typically modelled as a single-temperature grey-body emission (e.g. \citealt{capak15,Bouwens16,carniani2018alma,laporte19,big3drag,Bakx20,bakx:2021}, see also \citealt{Sommovigo:2021}), which is characterised mainly by
the dust temperature $T_{\rm d}$ and the dust mass $M_{\rm d}$.

Recently, the ALMA large program ALPINE (PI: Le Fèvre, \citealt{lefevre:2020,bethermin_alpine,FaisstALPINE}) provided us with the most abundant sample of FIR continuum detected galaxies at high-$z$ ($23$ at $z\sim 5$), featuring precise stellar masses $M_{\star}$ determinations thanks to the wealth of photometric data points available for individual targets. Interestingly, dust-to-stellar mass ratios as large as $M_{\rm d}/M_{\star} = 0.002-0.056$ have been inferred for ALPINE galaxies \citep{2021Pozzi}. Such large dust yields at these early epochs are in tension with theoretical predictions, adding to the so called \quotes{dust budget crisis} \citep[e.g. ][and references therein]{2014MNRAS.441.1040R}.

The favoured dust production mechanism at $z\geq 5$ is short lived supernovae (SNe, \citealt{2001:todini,lesniewska2019dust}), due to the stringent time constraints imposed by the age of the Universe combined with the young stellar populations in  galaxies\footnote{Asymptotic Giant Branch (AGB) stars are the main dust production sources at low-$z$. However, at $z \geq 5$ the galaxies ages are typically comparable to the lifetimes of AGB stars ($>150\ \mathrm{Myr}$), making their contribution to dust production likely sub-dominant (e.g. \citealt{Mancini2015,lesniewska2019dust,2019MNRAS.490..540L,2020burg,2020A&A...641A.168N,dayal:2022}).}. However, several works suggest that the dust yield produced in each SN event might be $\simlt 0.1\ \mathrm M_{\odot}$ after reverse shock processing  \citep{2016A&A...587A.157B,2019Matsuura,2020Slavin,dayal:2022}. This would imply very stringent limits on the expected dust-to-stellar mass ratios, $M_{\rm d}/M_{\star} \simlt 10^{-3}$, in tension with some high-$z$ measurements including the ALPINE ones \citep[see also][for higher-$z$ examples]{Tamura19,dayal:2022,Witstok22}. 

It is important to stress that these dust masses measurements are heavily dependent on the cold dust temperatures ($T_{\rm d} =25\ \mathrm{K}$ for ALPINE galaxies) \textit{assumed} in the FIR SED-fitting procedure. In fact, most FIR continuum observations at $z>4$ (typically in ALMA bands 6,7) probe a narrow wavelength range far from the emission peak, where different grey-body curves would deviate the most \citep[e.g.][]{Bouwens16,barisic2017dust,bowler2018obscured,big3drag,Tamura19}. Therefore, $T_{\rm d}$ is often fixed in the SED fitting procedure to reach convergence. 
As a result, all the quantities inferred from fitting, namely $M_{\rm d}$, the IR luminosity\footnote{$L_{\rm IR}$ is defined as the integrated continuum luminosity in the rest-frame wavelength range $8-1000\ \mathrm{\mu m}$.}, $L_{\rm IR}$, and the dust-obscured SFR ($\mathrm{SFR}_{\rm IR}/M_{\odot}yr^{-1} = 10^{-10} L_{\rm IR}/L_{\odot}$, \citealt{kennicutt1998}), are highly uncertain as they depend strongly on the assumed $T_{\rm d}$ ($\mathrm{SFR}_{\rm IR} \propto L_{\rm IR} \propto M_{\rm d} T_{\rm d}^6$ , see \citealt{behrens18,Liang19,sommovigo20} for a detailed discussion). 

Several theoretical works have suggested the presence of warmer dust ($T_{\rm d} \simgt 60\ \mathrm{K}$) in high-$z$ galaxies \citep{behrens18,Liang19,sommovigo20,pallottini:2022}, with temperatures as large as $\sim 100\ \mathrm{K}$ being reached in the most compact star-forming regions, Giant Molecular Clouds (GMCs, \citealt{behrens18,sommovigo20}). So far, warm dust temperatures $T_{\rm d}\sim 40-60\ \mathrm{K}$ have been measured in some of the few $z\geq5$ galaxies ($2$ of which are included in the ALPINE sample) for which multiple FIR data are available \citep{FaisstALPINE,bakx:2021,Witstok22}. Values as large as $T_{\rm d}>80\ \mathrm{K}$ have been measured in the only two galaxies detected in multiple ALMA bands at $z\sim 8$ \citep{laporte19,Bakx20}. 
Warmer dust temperatures have important implications. They reduce the requirements on $M_{\rm d}$ to produce the same observed FIR emission (reducing/eliminating the tension with dust production constraints from Supernovae, \citealt{behrens18,sommovigo20}). Warmer $T_{\rm d}$ values also imply larger obscured SFR fractions (up to $\sim 90\%$ as early as early as $z \sim 7$, e.g. \citealt{bakx:2021}), despite most of high-$z$ sources observed with ALMA are selected as UV-bright. 

Stacked SEDs analysis across a wide redshift range ($z=0-10$, e.g. \citealt{viero:2013,2014A&A...561A..86M,2015Bet,Schreiber18,bouwens:2021,2022:viero}), so far seem to confirm that on average dust is warmer at high-$z$, further suggesting the existence of a strong correlation between $T_{\rm d}$ and redshift. A consensus on the $T_{\rm d}-z$ evolution is yet to be reached, with the largest discrepancies arising at higher-$z$ end, where lesser and more uncertain data are available. 
In \cite{Sommovigo:2022} we proposed a physical model which motivates the increasing $T_{\rm d}-z$ trend with the decrease of the total gas depletion time $t_{\rm dep}=M_{\rm gas}/\mathrm{SFR}$ at high-$z$, due to the more vigorous cosmological accretion at earlier times. We show that $T_{\rm d} \propto t_{\rm dep}^{-1/6}$, implying a mild cosmic evolution as $T_{\rm d} \propto (1+z)^{0.42}$.

The purpose of this paper is to investigate the individual dust and FIR emission properties of the numerous $z\sim 5$ ALPINE galaxies. We are able to do so despite the availability of the single ALMA band-6 measurement thanks to a new method to derive $T_{\rm d}$ presented in \cite{Sommovigo:2021}. This method relies on combining the flux at rest-frame $1900\ \mathrm{GHz}$ ($158\ \mathrm{\mu m}$) with the overlying \CII\ emission-line. In particular, the \CII\ luminosity $L_{\rm CII}$ serves as a proxy for the dust mass, breaking the degeneracy between $M_{\rm d}$ and $T_{\rm d}$ in the FIR SED-fitting procedure. 

With our analysis we can infer the dust masses of ALPINE \CII\ and continuum detected sources without the need to rely on an assumption on $T_{\rm d}$. We can compare these new results with theoretical dust production constraints at $z\sim 5$. Moreover we can study individual galaxies $T_{\rm d}$, $L_{\rm IR}$ and $\mathrm{SFR}_{\rm IR}$, providing a complementary view to stacked FIR SEDs analysis \citep{bethermin_alpine}. 
Deriving ALPINE galaxies $T_{\rm d}$ also provides a fundamental anchoring to the $T_{\rm d}-z$ evolution recently extended in the Epoch of Reionization (EoR) thanks to the REBELS galaxies study at $z \sim 7$ (\citealt{Sommovigo:2022}, for details on the REBELS survey see \citealt{bouwens:2021}). ALPINE continuum-detected galaxies are almost $\sim 2$ times more numerous than REBELS ones, and span a wider SFR and $M_{\star}$ range (see Fig. \ref{MdMs}), thus constituting a less biased sample with respect to REBELS galaxies (which in turn have the advantage of probing higher-$z$, where empirical $T_{\rm d}-z$ relations differ the most). A consistent analysis and thorough comparison of both samples is crucial to solidify our understating of dust properties at high-$z$.


The paper\footnote{We assume a $\Lambda$CDM model with the following cosmological parameters: $\Omega_{\rm M} = 0.3075$, $\Omega_{\Lambda} = 1- \Omega_{\rm M}$, $\Omega_{\rm B} = 0.0486$,  $h=0.6774$, and $\sigma_8=0.81$. $\Omega_{\rm M}$, $\Omega_{\Lambda}$, $\Omega_{\rm B}$ are the total matter, vacuum, and baryonic densities, in units of the critical density; $h$ is the Hubble constant in units of $100\, {\rm km s}^{-1}$, and $\sigma_8$ is the late-time fluctuation amplitude parameter \citep{Planck16}.} is organised as follows. In Section \ref{method} we summarize the \citet[][]{Sommovigo:2021} method used to compute $T_{\rm d}$ and its application to ALPINE galaxies. We present our results on the dust temperatures and the $T_{\rm d}-z$ evolution in Sec. \ref{Td}. We then discuss the inferred dust masses and compare them with theoretical dust production constraints at $z\sim 5$ in Sec. \ref{Md}. In Sec. \ref{LIR and SFR} we discuss our estimates on the IR luminosities and obscured SFR fraction, and compare them with the SFR estimates previously obtained based on dust-corrected UV and optical data for ALPINE galaxies. We conclude with a brief summary of our results in Sec. \ref{summary}. 

\section{Method}\label{method}

In \cite{Sommovigo:2021} we proposed a novel method to derive the dust temperature in galaxies relying on a single ALMA measurement, by combining the continuum flux and the \CII\ line emission. We briefly summarize the method in the following.

We re-write the equation for the dust continuum flux $F_{\nu}$ observed against the CMB at rest-frame frequency $\nu=1900\ \mathrm{GHz}$ ($F_{\rm 1900}$) in a more compact form, yielding the following explicit expression for $T_{\rm d}$:
\begin{equation}\label{td1pto}
T_{\rm d} = \frac{T_{\rm 1900}}{\ln(1+f^{-1})}\,,
\end{equation}
where $T_{\rm 1900}= 91.86$ K is the temperature corresponding to the \CII\ transition energy at $1900$ GHz. The function $f$ is defined as: 
\begin{equation}\label{f}
f = [\exp(T_{\rm 1900}/T_{\rm CMB})-1)]^{-1} + A^{-1}\tilde F_{\rm 1900},
\end{equation}
where $T_{\rm CMB}$ is the CMB temperature at a given redshift. The non-dimensional continuum flux $\tilde F_{\rm 1900}$ and the constant
$A$ correspond to:
\begin{equation}\label{defs}
\begin{split}
    &\tilde F_{\rm 1900} = 0.98 \times 10^{-16} \left(\frac{F_{\rm 1900}}{\rm mJy} \right),\\[1mm]
    & A =  4.33 \times 10^{-24} \left[\frac{g(z)}{g(6)}\right] \left(\frac{M_{\rm d}}{M_\odot}\right)\,,\\[1mm]
\end{split}
\end{equation}
where $g(z) = {(1+z)}/{d_L^2}$ and $d_{\rm L}$ is the luminosity distance at redshift $z$.

We use the \CII\ luminosity, $L_{\rm CII}$, as a proxy for the total gas mass $M_{\rm g}$, and thus for $M_{\rm d}$ given a dust-to-gas ratio $D$. There is a consensus that $D$ scales linearly with the metallicity, $Z$, as $D=1/162\ (Z/Z_{\odot})$, with little scatter down to $Z\sim 0.1\ \mathrm{Z_{\odot}}$ \citep{james2002, 2007ApJ...657..810D,galliano2008,Leroy_2011,2014RR}. We adopt such scaling for $D$ since we do not expect metallicities $Z\ll 0.1\ Z_{\odot}$ in relatively evolved and massive ALPINE galaxies \citep{2022MNRAS.511.1303V}. We can then write the following expression for $M_{\rm d}$:
\begin{equation}\label{Mdust}
M_{\rm d} = D M_{\rm g} = D\, \alpha_{\rm CII} L_{\rm CII},
\end{equation}
where $\alpha_{\rm CII}$ is the \CII-to-total gas conversion factor.

An analytic expression for $\alpha_{\rm CII}$ is derived by parametrising in terms of empirical relations such as the Kennicutt–Schmidt relation \citep[][hereafter, KS]{kennicutt1998}, and the De Looze relation between $L_{\rm CII}-$SFR \citep{delooze14}, which has been shown to be valid for ALPINE galaxies by \cite{Schaerer_alpine}. This yields 
\begin{equation}\label{alfahz}
    \alpha_{\rm CII}=32.47\ \frac{y^2}{\kappa_{\rm s}^{5/7}}\  \Sigma_{\rm SFR}^{-0.29} \quad \frac{M_\odot}{L_\odot}\,,
\end{equation}
where $\Sigma_{\rm SFR}$ is the SFR surface density\footnote{$\Sigma_{\rm SFR}$ is in units of $\mathrm{[M_{\odot}yr^{-1}kpc^{-2}}$.} and $\kappa_{\rm s}$ is the \quotes{burstiness parameter} which quantifies deviations from the KS relation, $\kappa_s = \Sigma_{\rm SFR}/(10^{-12}\ \Sigma_{\rm gas}^{1.4})$ ($\kappa_{\rm s}>1$ for starbursts and $\kappa_{\rm s}<1$ for quiescent galaxies; see \citealt{ferraraCII,2019MNRAS.487.1689P,Vallini20}). The \CII-to-UV emission size ratio, $y=r_{\rm CII}/r_{\star}$, is introduced as there is growing evidence that at $z>4$ \CII\, emission extends further than star-forming regions \citep[$1.5 \simlt y \simlt 3$ at $z > 4$, see][]{carniani:2017oiii,carniani2018clumps,Carniani20,matthee2017alma,Matthee_2019,2019ApJ...887..107F,Fujimoto_alpine,2019Ryb,2020A&A...633A..90G}.

The method described here has been tested on a sample of $19$ local galaxies and $10$ galaxies at $z \simgt 4$ \citep{Sommovigo:2021,Sommovigo:2022,bakx:2021}. For all these galaxies\footnote{The $10$ high-$z$ galaxies test sample includes $4$ sources from the \cite{capak15} sample, followed-up in multiple ALMA bands by \cite{dusttemp2020}. Two of these $4$ sources (HZ4 and HZ6) are also included in the ALPINE sample, see the discussion in Sec. \ref{Td}.} multiple data points in the FIR SED are available, allowing us to compare our inferred dust temperatures with robust $T_{\rm d}$ estimates obtained with traditional SED fitting. We recovered consistent dust temperatures within $\pm 30\%$ uncertainty spanning the redshift range $z = 0-8.31$ as well as the temperature range $20\ \mathrm{K} \simlt T_{\rm d} \simlt 100\ \mathrm{K}$. We also tested our method on simulations, applying it to the $z \sim 6.7$ galaxy Zinnia (a.k.a. serra05:s46:h0643) from the SERRA simulation suite \citep{pallottini:2022}. Also in this case, we recover $T_{\rm d}$ in agreement with single-temperature grey body SED fitting performed at the frequencies corresponding to ALMA bands 6, 7, and 8.

\begin{figure*}
    \centering
    \includegraphics[width=0.9\linewidth]{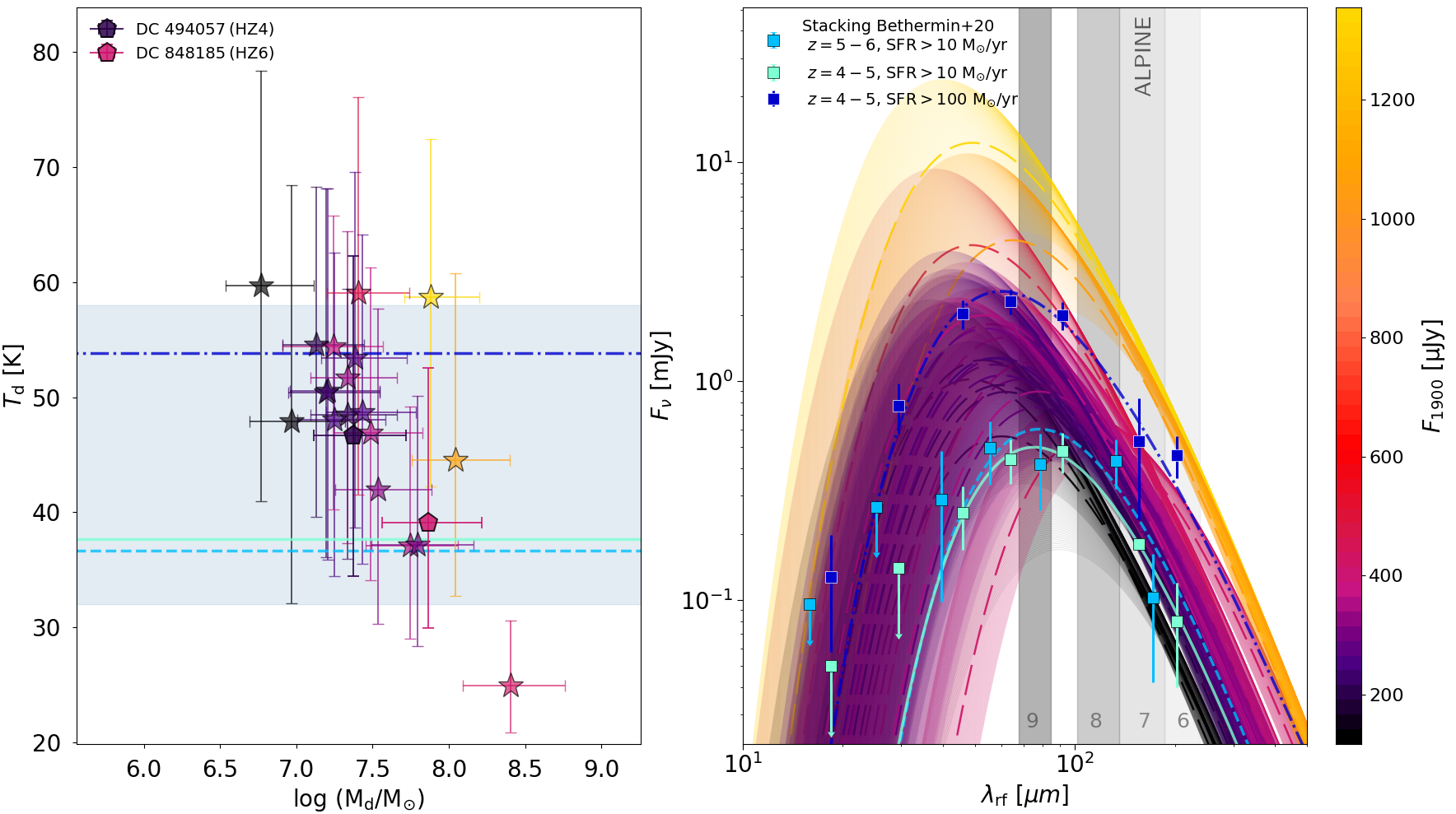}
    \caption{\textbf{Left Panel:} Dust temperature $T_{\rm d}$ and dust mass $M_{\rm d}$ derived for the ALPINE galaxies detected both in \CII\ and FIR continuum. The points are colour-coded according to their observed continuum flux $F_{1900}$. The two sources HZ4 and HZ6 for which multiple continuum data (and thus $T_{\rm d},M_{\rm d}$ estimates) are highlighted. The horizontal lines and blue shaded area show the $T_{\rm d}$ values and associated uncertainties, obtained by fitting the stacked data shown in the right panel with the same colors. \textbf{Right panel}: FIR SEDs obtained using the $(T_{\rm d},M_{\rm d})$ and associated uncertainty derived for ALPINE galaxies (shown in the left panel). The lines are colour-coded as in the left panel. The blue and green points show the stacked data by \citet{bethermin_alpine} at  $z = 4-6$. The three different colours correspond to different cuts in the redshift and SFR of the stacked sources: $z=4-5$ and $\mathrm{SFR}>10\ \mathrm{M_{\odot}/yr}$ (green), $z=4-5$ and $\mathrm{SFR}>100\ \mathrm{M_{\odot}/yr}$ (light blue), and $z=5-6$ and $\mathrm{SFR}>10\ \mathrm{M_{\odot}/yr}$ (darkblue). The grey shaded area marks the FIR wavelengths traced by ALMA bands $6-9$ (ALMA band $7$ is the one used in the ALPINE survey).}
    \label{fig:TdMd}
\end{figure*}

\subsection{Application to ALPINE galaxies}\label{applD}
We aim at deriving the dust masses and temperatures $(M_{\rm d}, T_{\rm d})$ of the $21$ ALPINE galaxies detected both in \CII\ and in continuum\footnote{We do not consider $2$ out of the $23$ \CII\ and continuum detected ALPINE sources, as their UV sizes, $r_{\star}$, and/or stellar masses $M_{\star}$ were not constrained.} at rest-frame $1900\ \mathrm{GHz}$.
The physical properties needed for the application of our method are
all constrained by observations\footnote{Measurements of $r_{\rm CII}$ are provided by \cite{Fujimoto_alpine} for all ALPINE sources except the galaxy VC 5101209780, which was left out due to its signal-to-noise ratio $\mathrm{SNR} < 5$. For this source, we measure the \CII\ radius in this work. We retrieve the ALMA continuum-subtracted, integrated \CII\ line map (from the ALPINE website, http://alpine.ipac.caltech.edu) and exploit the CASA task imfit software. We fit the \CII\ map with a 2D elliptical Gaussian model, and measure the beam-deconvolved FWHMs in arcseconds. We then infer $r_{\rm CII}$ from the mean of the beam-deconvolved major and minor axes. The result is given in Tab. \ref{tab:DATA_DET}. We have checked that this procedure gives $r_{\rm CII}$ estimates consistent to the ones provided by \cite{Fujimoto_alpine}, where available.} (see Tab. \ref{tab:DATA_DET} for all galaxies measured properties used in this work), with the exception of $(\kappa_{\rm s},Z)$. 

The burstiness parameter $\kappa_{\rm s}$ and metallicity $Z$ of ALPINE galaxies are unknown, thus a broad range of values is assumed for each quantity. 
Values as large as $\kappa_{\rm s} \simeq 100$ have been observed in star-forming galaxies both locally and at intermediate-redshift \citep[see e.g.][]{Daddi_2010}. By applying the \CII-emission model given in \cite{ferraraCII} on $12$ bright UV-selected galaxies at $z=6-9$, \cite{Vallini20,2021Vallini} found that $\kappa_{\rm s}$ values spanning the range $\sim 3-80$.
%
Based on these findings, for ALPINE galaxies we choose a random uniform distribution in the range $1 \simlt \kappa_{\rm s} \simlt 80$. 

For the metallicity, we assume a uniform random distribution of values in the range $0.3-1\ \mathrm{Z_{\odot}}$. This is based on numerical simulations results for galaxies at $z\simlt 6$ with similar stellar masses as ALPINE galaxies $10^9<M_{\star}/M_{\odot}<10^{11}$ \citep{Ma16, Torrey19}, and several observational studies which analyse FIR lines (such as \NII, \NIII, \CII, \OII\ and \OIII) to infer $Z$ in galaxies up to $z \sim 6-8$ \citep{sostpereira, big3drag,2019A&A...631A.167D,Tamura19,Vallini20,Bakx20,OIIImetal,2021arXiv211202115U}. In particular, \cite{2022MNRAS.511.1303V} infer an average metallicity $Z/Z_{\odot}\sim 0.5$ for 10 ALPINE galaxies based on their measured \OII-to-CII\ ratios. Better constraints on galaxies metallicities out to $z \sim 10$ will be available soon thanks to the 
James Webb Space Telescope (JWST) spectroscopic observations of rest-frame optical nebular lines (such as \Hb, \Ha, \NII, \OII\ and \OIII, \citealt{Maiolino_2019,Chevallard,2022arXiv220712375C}).


We can now compute the \CII-to-total gas conversion coefficient $\alpha_{\rm CII}$ for the ALPINE sources using eq. \ref{alfahz}. We find on average $\left< \alpha_{\rm CII} \right> = 8^{+3}_{-5}$, which is consistent with our previous findings for $z>4$ galaxies \citep{Sommovigo:2021,Sommovigo:2022}. Note that in local sources we find much larger conversion factors, up to  $\alpha_{\rm CII} \simlt 10^{3}$. This might suggest that for a given \CII\ luminosity high-$z$ galaxies have a lower gas content (or that for a given gas content they have higher \CII\ emission; for a detailed discussion see e.g. \citealt{ferraraCII}). For the comparison with the empirical \textit{molecular}-to-\CII\ conversion factor $\alpha_{\rm CII,mol}=\Sigma_{\rm H_2}/\Sigma_{\rm CII} = 31^{+31}_{-16}$ derived in \citet{zanita19} we refer to \citealt{Sommovigo:2022} (Sec. 3).\\

We can now estimate $M_{\rm d}$ and $T_{\rm d}$ for all the targets, and thus $L_{\rm IR}$ and $\mathrm{SFR_{\rm IR}}$. The results are summarised in Fig. \ref{fig:TdMd}, reported in Tab. \ref{tab:RES_DET}, and discussed in detail in the following Sections. 

\begin{figure*}
    \centering
    \includegraphics[width=0.9\linewidth]{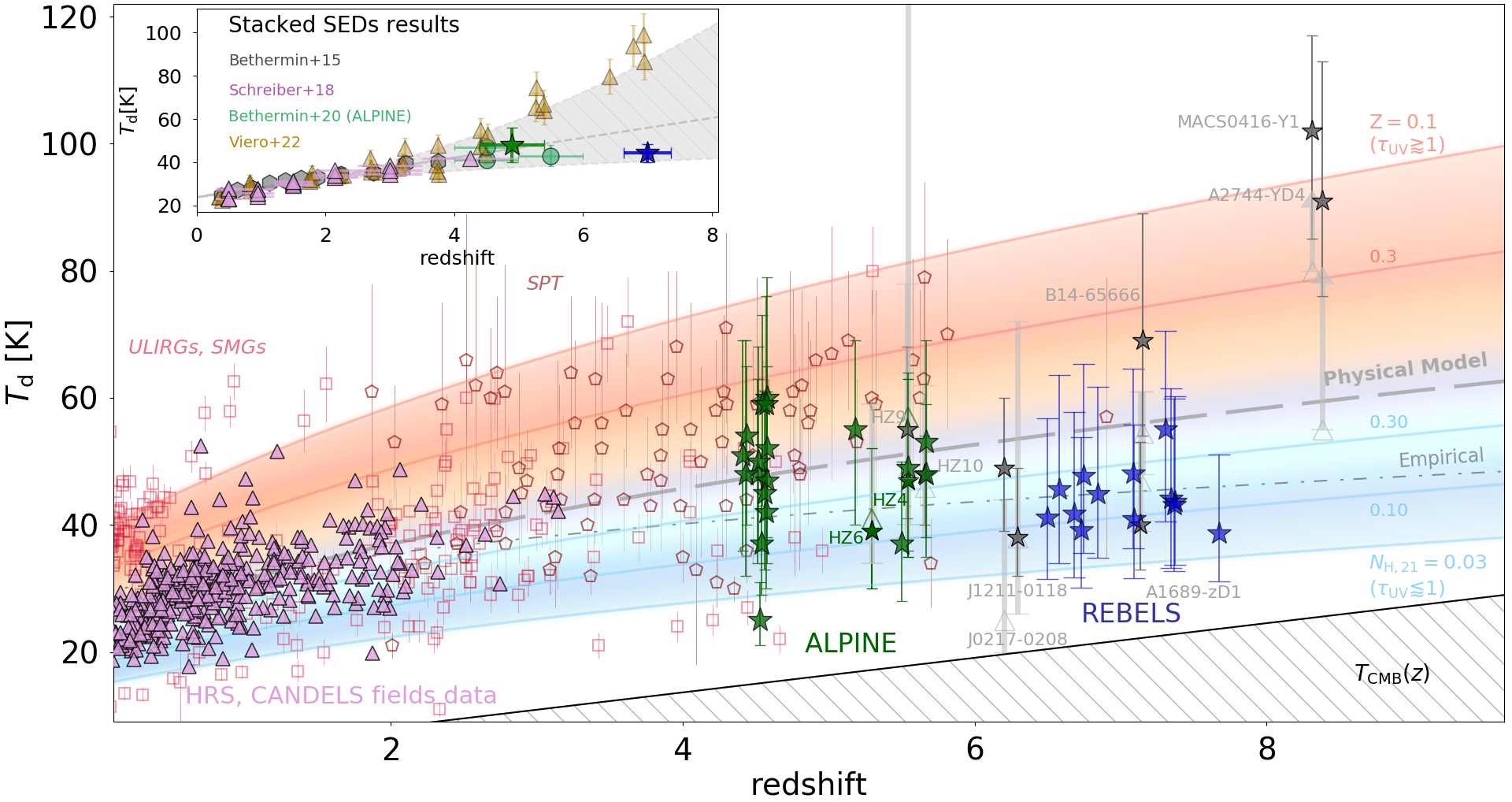}
    \caption{\textbf{Main panel:} Dust temperature $T_{\rm d}$ as a function of redshift for star-forming galaxies in the range $z=0-8$ \textit{(updated version of Fig. 2 in \citealt{Sommovigo:2022})}. The purple triangles represent the UV-to-IR normal star-forming galaxies detected in the HRS and CANDELS fields studied in \citet{Schreiber18}, which have comparable $M_{\star}$ to ALPINE and REBELS galaxies ($10^{9.5} \leq \mathrm{M_{\star}/M_{\odot}} \leq 10^{11.5}$). The maroon and red empty symbols represent respectively the sub-mm galaxies observed at $z\simlt 2$ \citep{2007Yang,Magdis_2014,Huang_2014,2005Chapman,Clements} and at $z=2-4$ (SPT sample, \citealt{ReuterSPT}). The grey points show the individual UV-selected galaxies at $z>5$ for which $T_{\rm d}$ estimates are available thanks to multiple FIR continuum observations. Both the $T_{\rm d}$ values obtained with our method and with traditional SED fitting are shown (respectively as stars and triangles; we note that in all cases the two estimates are consistent within $1-\sigma$, for further details see \citealt{Sommovigo:2022}). Finally, in green and blue we show the results obtained with our method for individual ALPINE and REBELS galaxies, respectively. The coloured region shows the $T_{\rm d}$-redshift evolution that we derive analytically in \citealt{Sommovigo:2022} (for an increasing effective UV optical depth $\tau_{\rm UV}$ from blue to red). We find that on average $T_{\rm d}$ raises with redshift due to the decreasing gas depletion time $t_{\rm dep}=M_{\rm g}/\mathrm{SFR}$ at higher-z, as $T_{\rm d}(z) \propto t_{\rm dep}^{-1/6}$. This dependence is shown by the grey dashed line for $t_{\rm dep}\propto (1+z)^{-2.5}$ -- derived from numerical simulations -- implying $T_{\rm d}\propto (1+z)^{0.4}$. For comparison, we also show the $T_{\rm d}-z$ relation we would predict based on the empirical evolution $t_{\rm dep,H_2}\propto (1+z)^{-1.5}$ inferred by \citealt{2020ARA&A..58..157T} for main-sequence galaxies (grey dotted-dashed line). \textbf{Inset panel}: $T_{\rm d}$ values obtained from stacked SEDs fitting in the redshift range $0\simlt z \simlt 8$ (grey, purple, green and yellow points, from \citealt{2015Bet,Schreiber18,bethermin_alpine,2022:viero}, respectively). The grey hatched area shows the different $T_{\rm d}-z$ trends empirically derived in these works, with the lower (upper) bound being set by \citealt{Liang19} (\citealt{2022:viero}). The linear $T_{\rm d}-z$ relation derived by \citealt{Schreiber18} is also shown. The green (blue) star corresponds to the average temperature derived here for ALPINE (REBELS) galaxies.}
    \label{Tdz}
\end{figure*}

\section{Dust temperatures}\label{Td}
We find the dust temperatures of ALPINE galaxies to lay within the wide range $25-60\ \mathrm{K}$, with an average value around $\left< T_{\rm d} \right> = 48 \pm 8\ \mathrm{K}$. The relative error associated to the individual sources $T_{\rm d}$ is around $\Delta T_{\rm d}/T_{\rm d} \sim 30\%$, with the dominant contributions to this uncertainty coming from the metallicity and burstiness parameter. Albeit large, this $\Delta T_{\rm d}/T_{\rm d}$ is comparable to that obtained from traditional SED fitting at similarly high-$z$ when multiple FIR data are available \citep[see e.g.][]{2020Harik,big3drag,bowler2018obscured}. 
Among the ALPINE targets there are two galaxies, DC 494057 and DC 848185 (HZ4 and HZ6 from \citealt{capak15}), for which ALMA bands 6,7, and 8 observations are available thanks to dedicated follow-up observations by \cite{dusttemp2020}. For both sources, the $T_{\rm d}$ values derived with our method are 
consistent with traditional SED fitting results within less than $\simlt 20\%$ uncertainty. 

In \cite{bethermin_alpine} the authors attempt at constraining the average $T_{\rm d}$ of ALPINE galaxies by fitting the stacked, single-dish FIR data from a photometric sample similar to ALPINE at $z=4-6$. 
The stacked targets in \cite{bethermin_alpine} are selected to resemble ALPINE galaxies in stellar mass, $M_{\star} > 3 \times 10^{10}\ \mathrm{M_{\odot}}$, and star formation rate, SFR$>10\ \mathrm{M_{\odot}/yr}$ (derived from optical and near-infrared SED fitting). They are divided in two redshift bins at $z=4-5$ (5749 sources) and $z=5-6$ (1883 sources). The comparison of these stacked data with the individual ALPINE galaxies SEDs derived with our method is shown in Fig. \ref{fig:TdMd}. We find that most (all but $3$) of the individual ALPINE galaxies FIR SEDs are bracketed by the stacked SEDs. In fact, by fitting the stacked SEDs with the same dust model adopted for the individual galaxies\footnote{We consider Milky Way-like dust, for which standard values for the dust opacity $\kappa_{\nu} = \kappa_{\star} (\nu/\nu_{\star})^{\beta_{\rm d}}$ are $(\kappa_{\star}, \nu_{\star}, \beta_{\rm d})$ = (10.41 ${\rm cm^2 g^{-1}}$, $1900\, {\rm GHz}$, $2.03$), from \citealt{weingartner2001dust,Draine03}.} we find $T_{\rm d}=37-54\ \mathrm{K}$, consistent (albeit slightly lower) with the range of dust temperatures obtained with our method. This consistency between SEDs of individual continuum-detected sources and stacked data is encouraging as stacking is widely used to extend star-forming galaxies IR emission studies out to $z \simgt 4$ \citep[see e.g.][]{2015Bet,Schreiber18,bethermin_alpine,2022:viero}. 


\subsection{Cosmic $T_{\rm d}$ evolution}\label{tdcosmic}
Recently, several works have alluded to the presence of a cosmic evolution of the dust temperature in star-forming galaxies \citep{2012Magdis,2013Magnelli,viero:2013,2015Bet,Schreiber18,FaisstALPINE,Liang19,Bouwens20,ReuterSPT}. In most of these studies, the average $T_{\rm d}$ at a given epoch is derived by fitting stacked data, including observations at wavelengths shorter than $\lambda \simlt 350\ \mathrm{\mu m}$ from the Herschel space observatory. Herschel was the only instrument probing these MIR-to-FIR wavelengths, which are crucial to obtain stringent constraints on $T_{\rm d}$. However, Herschel cannot detect individual galaxies at $z \simgt 2$ because of confusion \citep[e.g.][]{2012Magdis,2014A&A...561A..86M}, thus requiring to rely on stacking at higher-$z$. 

Interestingly, different works find discrepant results (see inset panel in Fig. \ref{Tdz}). Some works suggest a linearly increasing $T_{\rm d}-z$ trend based on stacked SEDs fitting results in the redshift range $z=0-5$ \citep{Schreiber18}, coupled with a few individual detections at $z\simgt 5$ \citep{Bouwens20}. Other works predict a much milder evolution \citep{2013Magnelli,viero:2013,ReuterSPT}; in particular \cite{Liang19} find a substantial flattening in the increase of $T_{\rm d}$ at $z \simgt 4$ based on SED fitting procedure applied to the FIRE simulations at $z=2-6$. This flattening in the $T_{\rm d}-z$ evolution is consistent with individual galaxies peak dust temperatures ($T_{\rm peak} \sim  2.9\times 10^3 (\lambda_{\rm peak}/\mathrm{\mu m})^{-1}$) measured at $z \sim 4-5$ by \cite{dusttemp2020}. However, $T_{\rm peak}$ can significantly differ from $T_{\rm d}$ depending on the adopted SED fitting function. In fact, \cite{dusttemp2020} finds $T_{\rm d} > T_{\rm peak}$ ($\delta T_{\rm d} \sim 13\ \mathrm{K}$ on average), consistently with our derivation (see Sec. \ref{Td}). It is worth mentioning that the only two $z\sim 8$ ALMA-detected sources host hot dust with temperatures $T_{\rm d}=90-100\ \mathrm{K}$ \citep[see also][]{behrens18,Bakx20,laporte19}, possibly questioning such scenario. We stress that for these $2$ galaxies we are able to uniquely derive $T_{\rm d}$ by combining the \CII\ luminosity information with the rest-frame $88\ \mathrm{\mu m}$ continuum flux\footnote{From traditional SED fitting only a lower limit on $T_{\rm d}>50,80\ \mathrm{K}$ is obtained \citep{Bakx20,laporte19}, see Appendix B in \citep{Sommovigo:2022}.}. These hot dust temperatures at $z\sim 8$ are consistent with the results by \cite{2022:viero}, where they exploit the recently released COSMOS2020
catalogue \citep{2022weaver} to extend stacked SEDs studies up to unprecedentedly high-$z$ ($z\sim 10$). The reliability of these stacking results at $z \gg 4$ is somewhat uncertain, as individuating low-$z$ interlopers and/or correcting for the bias towards the brightest sources becomes more challenging (see the discussion in \citealt{2022:viero}). We caution that \cite{2022:viero} find a nearly constant number of very massive sources ($\sim 60$ at $M_{\star}\simgt 10^{11}\ \mathrm{M_{\odot}}$) in their stacked bins at $z=3.5-4$ and $z=8-10$, which is in contrast with the predictions from the stellar mass function. In fact, extrapolating abundance matching results \citep{2019MNRAS.488.3143B} and observations \citep{Song_2016} at $z\leq 8$ and $M_{\star}\leq 3 \times 10^{10}\ \mathrm{M_{\odot}}$, we would expect a $> 2$ odm drop in the number density of $M_{\star}\simgt 10^{11}\ \mathrm{M_{\odot}}$ sources from $z=4$ to $z>8$. Semi-analytical models such as \code{DELPHI} \citep{dayal:2022} agree with this prediction, with a steep drop in the number density ($7$ odm) at the high-mass end ($M_{\star} = 4 \times 10^{10}\ \mathrm{M_{\odot}}$) from $z=4$ to $z=10$. Upcoming JWST observations will allow us to extend the census of massive systems at $z>8$, testing these predictions \citep{2022arXiv220709436C,2022arXiv220709434N,2022arXiv220712446L}.


Thanks to our method, for the first time we constrain $T_{\rm d}$ in a large number of sources ($40$) at $z\simgt 4$, thus adding fundamental and highly complementary information from individual galaxies analysis to stacked SED results. At $z=4.9$, the mean redshift of \CII\ and continuum detected ALPINE galaxies, we find an average $\left<  T_{\rm d} \right> = 48 \pm 8\ \mathrm{K}$, whereas $ \left< T_{\rm d} \right> = 44 \pm 4\ \mathrm{K}$ for the $13$ \CII\ and continuum-detected REBELS galaxies at $z\sim 7$ \citep{bouwens:2021,inami:2022}. We note that this value for REBELS galaxies $ \left<  T_{\rm d} \right>$ is slightly lower than the one reported in \cite{Sommovigo:2022} ($T_{\rm d} =47 \pm 7\ \mathrm{K}$). We have updated the stellar masses of REBELS galaxies to the latest values by \cite{2022Topping}, obtained by assuming a non-parametric star-formation history (SFH, instead of constant SF). A non-parametric SFH results in an increase in $M_{\star}$ up to one order of magnitude for galaxies with large specific SFR\footnote{This increase in $M_{\star}$ is due to the presence of a significant old stellar population that is out-shined by the recent star formation (SF) burst}. Larger $M_{\star}$ imply slightly larger dust $M_{\rm d}$ and lower $T_{\rm d}$ to reproduce the same observed $F_{1900}$.

The comparable average $T_{\rm d}$ in ALPINE and REBELS galaxies, whose main difference is the redshift of the sources\footnote{REBELS galaxies stellar masses and SFRs are similar -albeit spanning a narrower range- to that of ALPINE galaxies, being $9 \simlt \log( M_{\star}/M_{\odot}) \simlt 10$ and $20 \simlt \mathrm{SFR/M_{\odot} yr^{-1}} \simlt 200$.}, questions the validity of the simple, linearly increasing $T_{\rm d}-z$ trend suggested by \cite{Schreiber18,Bouwens20} (predicting an increase from $T_{\rm d}\sim 46\ \mathrm{K}$ at $z=4.9$ to $56\ \mathrm{K}$ at $z = 7$). On the other hand, the flattening in the $T_{\rm d}-z$ trend at $z>4$ inferred by \citet{2013Magnelli,Liang19} seems too extreme (they predict colder temperatures, $T_{\rm d} \sim 37\ \mathrm{K}$ at $z=4.9$ and $T_{\rm d} \sim 39\ \mathrm{K}$ at $z=7$). 
The largest discrepancy is with the preliminary results by  \cite{2022:viero}, whose best fitting $T_{\rm d}-z$ relation features a sharp increase in $T_{\rm d}$ at $z\simgt 5$ reaching $T_{\rm d}\sim 87\ \mathrm{K}$ at $z=7$ (almost $\times 2$ higher than the average dust temperature we find in REBELS galaxies at the same redshift).
A possible caveat is that both ALPINE and REBELS sources are UV-selected as the brightest sources at their respective redshift, thus constituting a biased sample, possibly skewed toward colder dust temperatures \citep[e.g.][]{chen2022}. JWST will allow us to probe relatively UV-faint galaxies also at $z\geq 5$; by following-up JWST observations with ALMA, we will investigate also UV-faint high-$z$ galaxies dust properties, possibly reducing the current observational bias. 

In \cite{Sommovigo:2022} we produced a model aimed at physically motivating the cosmic evolution of the dust temperature. Assuming FIR and UV emission to be co-spatial, from simple conservation of energy argument, we show that $T_{\rm d}$ anti-correlates with the \textit{total} gas depletion time as $T_{\rm d} \propto t_{\rm dep}^{-1/6}$. The increase of the cosmological accretion rate at early times \citep{Fakhouri10,Dekel13,2015DM}, results in high-$z$ galaxies being more efficiently star forming, thus implying shorter $t_{\rm dep}$ at high-$z$. As a result, we predict mild increase of $T_{\rm d}$ with redshift as:
\begin{equation}
    T_{\rm d} \propto t_{\rm dep}^{-1/6} \approx (1+z)^{0.42}.
\end{equation} 
where the adopted $t_{\rm dep}-z$ evolution ($t_{\rm dep} \propto (1+z)^{-2.5}$) is taken from numerical simulations \citep{Fakhouri10,Dekel13,2015DM} due to the lack of observational constraints on the atomic gas content of high-$z$ galaxies. The \textit{molecular} gas depletion time, $t_{\rm dep,H_2}=M_{\rm H_2}/\mathrm{SFR}$, has been indirectly studied up to $z\sim 6$ relying on CO and dust observations \citep[ e.g.][]{walter2020evolution,2020ARA&A..58..157T,2020dessauges}. For main-sequence galaxies, different works consistently infer an evolution of $t_{\rm dep,H_2}$ with redshift around $t_{\rm dep,H_2} \propto (1+z)^{-1.5}$ \citep{2020ARA&A..58..157T}. Adopting such empirical $t_{\rm dep,H_2}(z)$ relation, would imply an milder - but still significant- cosmic $T_{\rm d}$ evolution $T_{\rm d} \propto (1+z)^{0.25}$. This is also shown in Fig. \ref{Tdz}.

On top of the $T_{\rm d}-z$ trend, \cite{Sommovigo:2022} showed that the scatter in the measured $T_{\rm d}$ values at a given redshift can be explained by the variation of a few key individual galaxies properties, namely the optical depth $\tau_{\rm UV}$, metallicity and column density $N_{\rm H}\sim 10^{21} \tau_{\rm UV}/Z\ \mathrm{cm^{-2}}$. 
The results derived in this paper for individual ALPINE galaxies are consistent with the predictions from our physical model, as shown in Fig. \ref{Tdz}. Due to the different redshifts of the ALPINE and REBELS samples, we expected a minor mean temperature variation $|\left< T_{\rm d,REB} \right> - \left< T_{\rm d,ALP} \right>| /\left< T_{\rm d,ALP} \right> \sim 14\%$. However, such variation is comparable to the $1-\sigma$ error associated to $\left< T_{\rm d} \right>$ in the two samples. 

The slightly warmer $T_{\rm d}$ of ALPINE galaxies with respect to REBELS galaxies could be explained by larger UV optical depths, implying a more efficient dust heating. This prediction is qualitatively confirmed by the measured average UV slopes $\left< \beta_{\rm UV} \right>$ for the two samples. In fact, $\left< \beta_{\rm UV} \right>$ is redder in ALPINE galaxies ($-1.4 \pm 0.5$) than in REBELS galaxies ($-1.8 \pm 0.3$), implying a larger dust-obscuration (as a reference, we remind the intrinsic UV slope\footnote{The quoted intrinsic UV slope is obtained from \code{STARBUST99} \citep{starburst99}, assuming continuous star formation, Salpeter IMF $1-100\ \mathrm{M_{\odot}}$, metallicity $Z = 1/3\ \mathrm{Z_{\odot}}$, and stellar age $150\ \mathrm{Myr}$.} amounts to $\beta_{\rm int}= -2.406$, \citealt{ferrara:2022}). Our prediction of warmer dust temperatures in more UV-obscured systems is also confirmed by the comparison with sub-millimeter galaxies studies at $z\simlt 6$ (see Fig. \ref{Tdz}). In fact, on average sub-mm galaxies host warmer dust than UV-trasparent galaxies \citep[see also][]{dusttemp2020}, reaching values around $T_{\rm d}\sim 40\ \mathrm{K}$ in the local Universe \citep{2007Yang,Magdis_2014,Huang_2014,2005Chapman,Clements}, and $T_{\rm d}\sim 60\ \mathrm{K}$ at $z \sim 5$ \citep{ReuterSPT}\footnote{We caution that the SED-fitting procedure used in \cite{ReuterSPT} for SPT galaxies is different with respect to the one adopted here, which assumes an optically thin grey-body emission. Nevertheless, when testing our method on the only SPT galaxy with available metallicity measurements (SPT 0418-47), we found consistent results with \cite{ReuterSPT} within $1-\sigma$ \citep{Sommovigo:2021}.}. 


\section{Dust masses}\label{Md}
We derive the dust masses for our sample of the ALPINE galaxies finding them to vary within the range $ 6.77 < \log (M_{\rm d}/M_{\odot}) < 8.41$, with the average value being $\left< \log (M_{\rm d}/M_{\odot}) \right> = 7.47 \pm 0.37$. This value is a factor $\sim 7$ lower than the values reported in \cite{2021Pozzi} for these same galaxies. 
\begin{figure}
    \centering
    \includegraphics[width=1.0\linewidth]{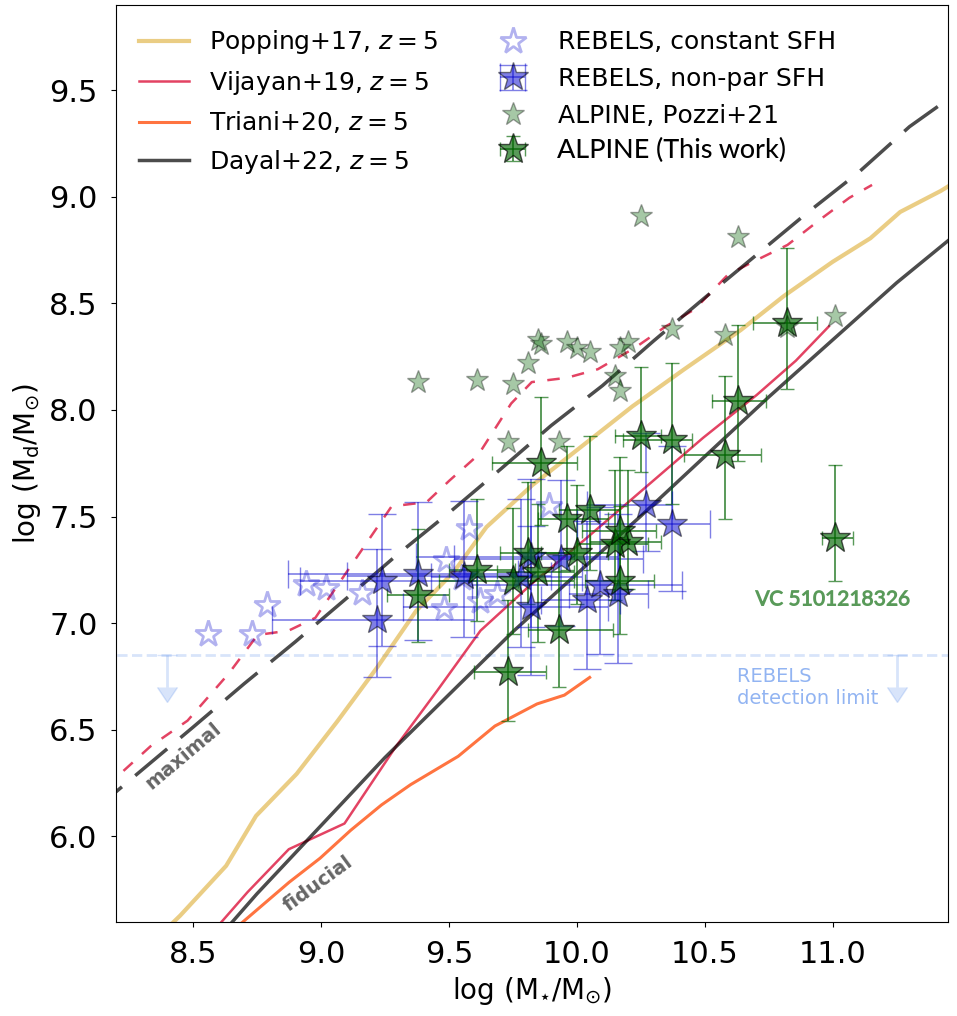}
    \caption{Dust mass $M_{\rm d}$ as a function of stellar mass $M_{\star}$. The solid green stars show the $M_{\rm d}$ values that we derive for ALPINE galaxies, whereas the transparent ones were obtained by \citet{2021Pozzi} assuming $T_{\rm d}=25\ \mathrm{K}$. The blue stars (empty stars) represent REBELS galaxies, where $M_{\star}$ is inferred assuming a non-parametric SFH (constant SF, \citealt{2022Topping}). The solid lines show the fiducial predictions at $z\sim 5$ from semi-analytical models such as \code{DELPHI} \citep[black,][]{delphi:2014,dayal:2022}, Santa Cruz \citep[yellow,][]{10.1093/mnras/stx1545}, \code{Dusty Sage} \citep[orange,][]{2020MNRAS.493.2490T}, and \code{L-Galaxies} \citep[red,][]{10.1093/mnras/stz1948}. The dashed lines show the maximal predictions of the corresponding models, assuming no dust destruction or ejection, and saturated grain growth. We find that all but one ALPINE galaxies are consistent with theoretical predictions. This is not true for higher-$z$ REBELS galaxies (particularly under the assumption of constant SF), whose $M_{\rm d}-M_{\star}$ relation appears flat; possible explanations are discussed in the text (see also \citealt{dayal:2022}). The dashed blue line shows the $M_{\rm d}$ detection threshold for the REBELS program assuming a non-detection flux limit of $42\ \mathrm{\mu Jy}$ and $T_{\rm d}=48\ \mathrm{K}$.}
    \label{MdMs}
\end{figure}
\cite{2021Pozzi} \textit{assume} a very cold dust temperature of $25\ \mathrm{K}$, to derive the dust masses from the continuum fluxes $F_{\rm 1900}$. They state that this value (which is close to $T_{\rm CMB}(z=6)$) should correspond to the mass-weighted dust temperature of typical $z\sim 6$ galaxies. However, by analysing an average simulated galaxy at $z\sim 6$ from the SERRA simulation suite \citep{2019MNRAS.487.1689P,pallottini:2022}, in \cite{Sommovigo:2021} we found that the mass-weighted dust temperature and $T_{\rm d}$ derived from fitting the simulated spectra (using mock continuum observations in ALMA band 6,7, and 8 or our method) actually correspond, and are $\gg T_{\rm CMB}$ ($T_{\rm d}\sim 60\ \mathrm{K}$, see also \citealt{pallottini:2022} for an extended discussion). The temperatures that we infer for ALPINE galaxies are similarly warm, $T_{\rm d} \sim 48\ \mathrm{K}$, with a single exception represented by the galaxy VC 5180966608 for which we infer $T_{\rm d} = 25\ \mathrm{K}$. This cold dust results from the peculiar, and likely unreliable, \CII-to-UV size ratio measured for this galaxy. In fact, from eq. \ref{Mdust}-\ref{alfahz} it follows that $M_{\rm d} \propto y^{2}=(r_{\rm CII}/r_{\star})^2$, resulting in a much larger dust mass for VC 5180966608, whose $y=10$, with respect to the other galaxies in the ALPINE sample, where on average $\left< y \right> \sim 3$. This massive dust content ($\log (M_{\rm d}/M_{\odot}) = 8.41$), coupled with a relatively low continuum flux ($F_{\rm 1900}= 462\ \mathrm{\mu Jy}$), results in the exceptionally cold dust temperature derived for this galaxy. However, the \CII\ size of VC 5180966608 is flagged as unreliable by \cite{Fujimoto_alpine} due to its complicated morphology. In fact, in the latest analysis by \cite{2021A&A...653A.111R} VC 5180966608 is classified as a merger; treating it as a single source might have lead us to misinterpreting its properties. Upcoming deeper ALMA \CII\ observations and higher resolution FIR continuum data from the CRISTAL large program (PI: Herrera-Camus) will allow us to further investigate this hypothesis.  

The lower dust masses that we infer for ALPINE galaxies have important implications in terms of the comparison with theoretical dust production constraints at $z\sim 5$. We discuss this in detail in the following Section. 

\subsection{Dust production at $z \simgt 5$}
We begin by computing the dust yield $y_{\rm d}$ per SN which would be required to produce the dust masses derived in this work for ALPINE galaxies. This is $y_{\rm d}/M_{\odot} = M_{\rm d} /(\nu_{\rm SN} M_{\star})$, where $\nu_{\rm SN} = (53\ \mathrm{M_{\odot}})^{-1}$ is the rate of SNe per solar mass of stars formed assuming a Salpeter $1-100\ \mathrm{M_{\odot}}$ IMF \citep{10.1046/j.1365-8711.2000.03209.x}. We find that on average $\left< y_{\rm d} \right> = 0.15 \pm 0.09\ \mathrm{M_{\odot}}$, which is consistent with the SN dust production constraints by \cite{lesniewska2019dust}. However, other works suggest that the dust yield spared in a SN blast is as low as $\simlt 0.1 \mathrm{M_{\odot}}$ \citep{2016A&A...587A.157B,Matsuura2019,Slavin2020}. 

In order to investigate dust production further, in Fig.\ref{MdMs} we compare the dust-to-stellar mass relation that we find for ALPINE galaxies with predictions from semi-analytical dust production models at $z\sim 5$. We include the \code{DELPHI} model \citep{delphi:2014,dayal:2022}, the Santa Cruz model \citep{10.1093/mnras/stx1545}, \code{Dusty Sage} \citep{2020MNRAS.493.2490T}, and \code{L-Galaxies} \citep{10.1093/mnras/stz1948}, which (mostly) cover the ALPINE stellar mass range $M_{\star}=10^9-10^{11}\ \mathrm{M_{\odot}}$.
All these models include varying prescriptions for gas cooling, star formation, SN feedback, chemical enrichment and key dust processes, namely dust formation, astration, destruction in SNe shocks, ejection in outflows, and grain growth. 
The \code{DELPHI} model does not include the contribution of AGB stars to dust production, which is likely sub-dominant due to the long timescales required for such dust production mechanism ($\simgt 150\ \mathrm{Myr}$) and the conflicting young stellar ages of $z\simgt 5$ galaxies (e.g. \citealt{lesniewska2019dust,2019MNRAS.490..540L,2020A&A...641A.168N,dayal:2022}). 

Our results are consistent (within $1-\sigma$) with \code{DELPHI} and \code{L-Galaxies} theoretical predictions, with the Santa Cruz model favouring slightly larger dust masses (for nearly $50\%$ of the sample). The slope of the $M_{\rm d}-M_{\star}$ relation is consistent with our results in all the four models. therefore, it is not necessary to invoke unphysical scenarios with no dust destruction or ejection, and saturated grain growth. Moreover, dust production from SNe described by the \code{DELPHI} model is able to reproduce the inferred dust masses for most of the ALPINE galaxies ($72\%$ within $1-\sigma$ and the remaining ones within $1.5-\sigma$), confirming that the contribution to dust production from AGB stars in ALPINE galaxies is subdominant. 
The most discrepant galaxy is VC 5101218326, for which we predict a surprisingly low dust mass $M_{\rm d}=10^{7.4}\ \mathrm{M_{\odot}}$, roughly one order of magnitude below theoretical predictions given the galaxy large stellar mass $M_{\star}=10^{11}\ \mathrm{M_{\odot}}$. 
This source is one of the two most peculiar sources in the ALPINE sample in terms of its dust properties and will be discussed in detail in the following Section. 

Finally, we compare our results for the dust-to-stellar mass ratios in ALPINE galaxies, with the ones obtained for the REBELS sources with the same method \citep{Sommovigo:2022,dayal:2022}. By using the larger $M_{\star}$ values derived for REBELS galaxies assuming a non-parametric SFH instead of constant SF \citep{2022Topping}, the discrepancy with theoretical models is reduced (none of the sources exceeds the maximal dust production constraints). In fact, for less massive objects ($\log (M_{\star}/M_{\odot})\leq 10^{9.5}$) the increase in $M_{\star}$ can be as large as $1\ \mathrm{dex}$, while $M_{\rm d}$ is marginally affected ($30\%$ variation). 
However, differently from ALPINE galaxies, in the higher-$z$ REBELS sample we do not recover the $M_{\rm d}-M_{\star}$ correlation predicted by analytical models (independently from the assumed SFH). In fact, REBELS galaxies dust masses appear to be independent from their stellar masses. ALPINE galaxies are more massive than REBELS objects, and overall cover a larger range of $M_{\star}$ values. This might indicate that flatness of the $M_{\rm d}-M_{\star}$ trend found for REBELS galaxies largely depends on an observational bias due the limited range of stellar masses probed by this survey\footnote{In Fig. \ref{MdMs} we show the $M_{\rm d}$ detection threshold for the REBELS program assuming a non-detection flux limit of $42\ \mathrm{\mu Jy}$ \citealt{inami:2022} and the average dust temperature $T_{\rm d}=47\ \pm 7\ \mathrm{K}$.}. Further ALMA observations probing the dust content of galaxies in a wider stellar mass range both at $z=7$ and $z=5$, will confirm whether we are witnessing an evolution in the dust-to-stellar mass relation between these two epochs. This could imply that also the mass-metallicity relation (to whom our dust-to-stellar mass relation is directly connected as we assume $D \propto Z$) breaks down at $z>6$. 


\section{IR luminosities and total SFR}\label{LIR and SFR} 
We compute the IR luminosities using the following relation \citep{ferrara:2022,Sommovigo:2022}:
\begin{equation}\label{LIR}
{L_{\rm IR}}=\left( \frac{M_{\rm d}}{M_{\odot}} \right)\ \left( \frac{T_{\rm d}}{8.5\, \rm K} \right)^{6.03} L_{\odot}.    
\end{equation}
valid for the Milky Way (MW) dust model adopted here.
We find that ALPINE galaxies IR luminosities vary in the range  $1.7 \times 10^{11}\ \mathrm{L_{\odot}} \simlt L_{\rm IR}\simlt 8.7 \times 10^{12}\ \mathrm{L_{\odot}}$. Among the $21$ galaxies analyzed here, as many as $8$ have IR luminosities comparable to Ultra-Luminous InfraRed Galaxies\footnote{For comparison, only one ULIRG-like galaxy, REBELS-25, is found among the $13$ REBELS galaxies (\citealt{Sommovigo:2022,inami:2022}, see also Algera in prep. for a detailed analysis of the source).} (ULIRGs, i.e. $L_{\rm IR}>10^{12}\ \mathrm{L_{\odot}}$, see \citealt{Lonsdale06}). This finding is quite surprising as ALPINE galaxies are selected as UV-brightest sources in the given redshift range $z=4-6$. Our results suggest that a large fraction of their star formation is dust-obscured, which is consistent with stacking results by \cite{2020A&A...643A...4F} (suggesting that on average $45\%$ of SFR is obscured at $z=5-6$).

The inferred IR luminosities correspond to obscured SFRs in the range $\mathrm{SFR_{\rm IR}} \sim 17-878\ \mathrm{M_{\odot}/yr}$ (assuming the conversion factor given in Table \ref{tab:RES_DET}). The largest value for $\mathrm{SFR_{\rm IR}}=878\ \mathrm{M_{\odot}/yr}$ is found in the galaxy DC 873756, the least UV-bright galaxy among the FIR continuum detected ALPINE sources. Its UV magnitude ($M_{\rm UV}=-20.9$) corresponds to a monochromatic luminosity at $1500\ \angstrom$ around $L_{\rm UV}=10^{10.7}\ \mathrm{L_{\odot}}$, implying an anattenuated\footnote{$\mathrm{SFR}_{\rm UV}=L_{\rm UV}/{\cal K}_{1500}$, where the conversion coefficient ${\cal K}_{1500} =1.174\times 10^{10}\ \mathrm{L_{\odot}/(M_{\odot}yr^{-1})}$ is taken from \cite{ferrara:2022}.} $\mathrm{SFR_{\rm UV}=4\ \mathrm{M_{\odot}/yr}}$. Compared to its surprisingly large $\mathrm{SFR}_{\rm IR}$, this implies that more than $99\%$ of the SFR in DC 873756 is obscured\footnote{Following the same procedure, we infer obscured SFR fractions in the range $\sim 60-90\%$ for the remaining ALPINE galaxies.}. These results is highly in contrast with UV-to-optical SED-fitting results by \cite{FaisstALPINE}, who derive a total, dust-corrected $\mathrm{SFR_{\rm SED}}=5^{+10}_{-2}\ \mathrm{M_{\odot}/yr}$. Using only UV-to-optical data the total SFR of DC 873756 was underestimated by more than two orders of magnitude\footnote{We note that a milder correction to account for the obscured SFR fraction was already applied by \cite{Schaerer_alpine} to some ALPINE galaxies; due to the lack of constraints on individual galaxies IR luminosities, they relied on the dust temperature derived by \cite{bethermin_alpine} from stacking.}. 

\begin{figure}
    \centering
    \includegraphics[width=1.0\linewidth]{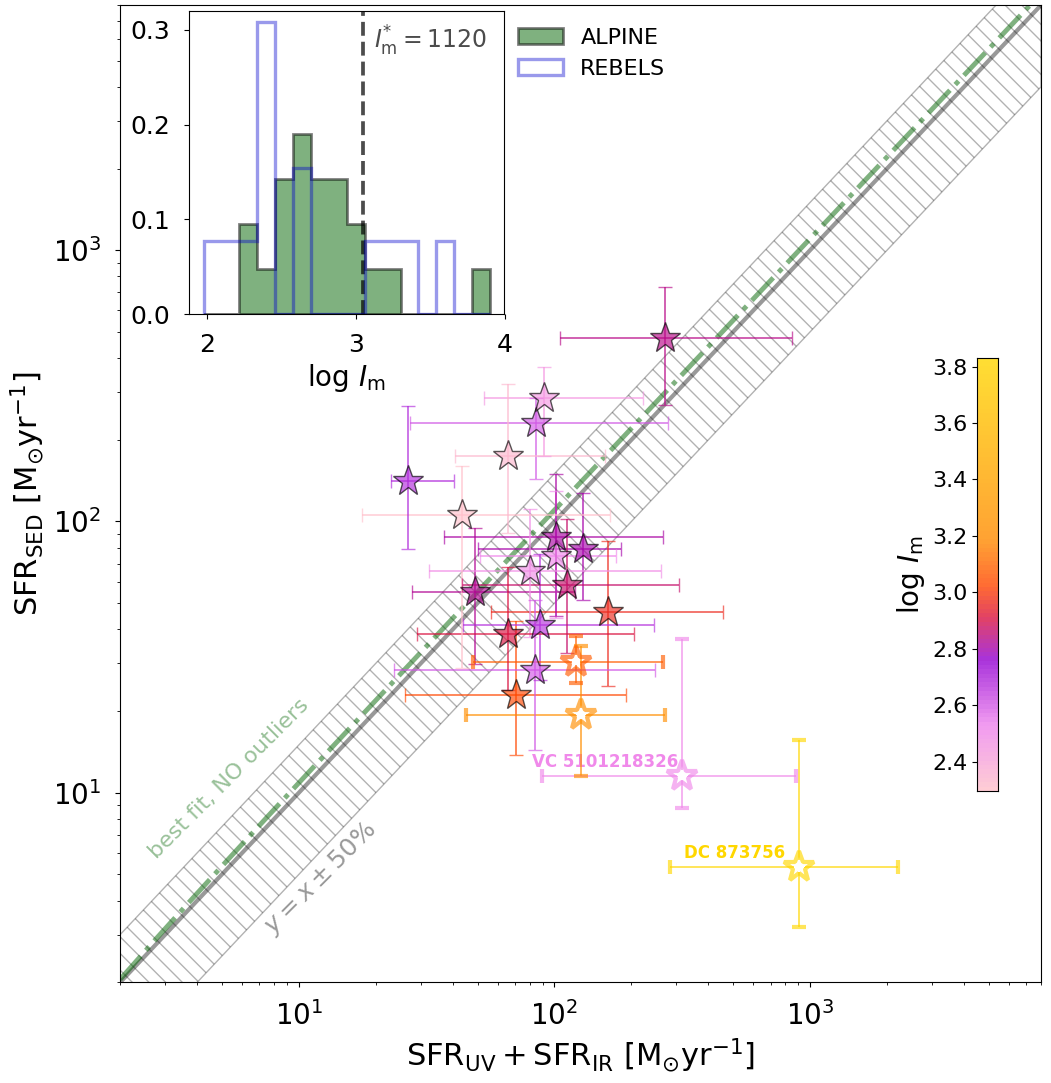}
    \caption{\textbf{Main panel}: Total SFR inferred for ALPINE galaxies through dust-corrected optical-to-UV SED fitting, $\mathrm{SFR}_{\rm SED}$ \citep{FaisstALPINE} vs. that obtained from the UV and IR luminosities multiplied by the corresponding calibration factors, $\mathrm{SFR_{\rm UV}+ SFR_{\rm IR}}$. The dashed area shows the area around the bisector $\pm 50\%$ uncertainty. The light green symbols represent the most peculiar sources in the sample, for which $\mathrm{SFR_{\rm IR}> 100\ \mathrm{M_{\odot}/yr}}$, largely deviating from the SFR deduced from optical-to-UV data $\mathrm{SFR_{\rm SED}< 30\ \mathrm{M_{\odot}/yr}}$. The green dotted dashed line shows slope of the best fitting linear relation obtained when excluding these outliers, consistent with the bisector. \textbf{Inset panel:} Distribution of the molecular index $I_{\rm m}=(F_{\rm 1900}/F_{\rm UV})/(\beta_{\rm UV}-\beta_{\rm int})$ among the ALPINE continuum detected galaxies. The vertical dashed line shows the upper limit $I_{\rm m}^{*}=1120$ obtained in a single phase ISM model, where UV and IR emission are cospatial. This limit is exceeded in $3$ among the $21$ ALPINE galaxies considered here, which correspond to the outliers in the main plot (but one). We speculate that these peculiar galaxies have spatially segregated UV and IR emission.   }
    \label{SFRcomp}
\end{figure}

In order to understand the nature of this discrepancy and whether DC 873756 represents an isolated case, in Fig. \ref{SFRcomp} we compare the SFR derived from UV-to-optical SED-fitting with $\mathrm{SFR_{\rm UV}}+\mathrm{SFR_{\rm IR}}$ for all the ALPINE sources considered here. We find that in most cases ($80\%$) the two methods give consistent results within $1-\sigma$. The uncertainties on $\mathrm{SFR_{\rm IR}}$ are very large (see Tab. \ref{tab:RES_DET}) due to the strong dependence of this quantity on $T_{\rm d}$ ($\mathrm{SFR_{\rm IR}} \propto L_{\rm IR} \propto T_{\rm d}^6$), which is only constrained within $\pm 10\ \mathrm{K}$. There are $4$ outliers (DC 539609, VC 5100969402, DC 873756, VC 5101218326); for these sources the ratio\footnote{This discrepancy is larger than what can be explained by the different assumptions on the IMF in the two derivations (here we assume a Salpeter $1-100\ \mathrm{M_{\odot}}$, whereas \citealt{FaisstALPINE} adopt a Chabrier.} between the UV-to-optical SED-derived SFR and that obtained from our method is $<1/3$. In all these sources strong FIR continuum emission ($\mathrm{SFR_{\rm IR}}>100\ \mathrm{M_{\odot}/yr}$) coexists with blue UV slopes $\beta_{\rm UV}<-1$, which would in contrast suggest low dust obscuration. These $4$ galaxies are highlighted in Fig. \ref{SFRcomp}, with the two most extreme cases being the galaxies DC 873756 and VC 5101218326. We note that if we do not include these two sources in the $\mathrm{SFR}_{\rm SED}$ vs. $\mathrm{SFR}_{\rm UV}+\mathrm{SFR}_{\rm IR}$ linear fit we recover a slope which is perfectly consistent with the bisector $1.1 \pm 0.4$ (albeit the large scatter), whereas if we include them we find $-0.1 \pm 0.1$.

One possible scenario to explain these peculiar galaxies is that they host Active Galactic Nuclei (AGN). Indeed, \citep{dimascia:inprep} show that AGN can emit significantly not only at MIR, but also at FIR wavelengths. Thus, not accounting for AGN contribution, results in an overestimation of the host galaxy $L_{\rm IR}$ and obscured $\mathrm{SFR_{\rm IR}}$. However, this scenario seems unlikely based on these ALPINE galaxies optical-to-UV spectra, which do not show any peculiar feature with respect to the other sources in the sample. Nevertheless, whether AGN could contribute to the FIR emission of the most massive ALPINE galaxies such as VC 5101218326 (whose $M_{\star} = 10^{11}\ \mathrm{M_{\odot}}$ is the largest among all ALPINE galaxies) is still an open question (Barchiesi et al. in prep., Shen at al. in prep, \citealt{2022arXiv220603510F}). 

A clue for the interpretation of these outliers comes from computing their molecular index \citep{ferrara:2022}:
\begin{equation}\label{Im}
    I_{\rm m} = \frac{(F_{1900}/F_{\rm UV})}{(\beta_{\rm UV}-\beta_{\rm int})}
\end{equation}
where $\beta_{\rm int}=-2.406$ for the MW. Assuming the interstellar medium (ISM) to be described by a single zone model, where dust and stars are uniformly mixed\footnote{And UV and IR emission are co-spatial}, \cite{ferrara:2022} obtains the following analytical expression for $I_{\rm m}\sim 7062\ x e^{-3x^{1/6}}$, where $x=Z t_{\rm dep}/(\beta-\beta_{\rm int})$. This expression has a maximum $I_{\rm m}^* \simeq 1120$ (located at $x=64$). Indeed, $F_{1900}$ and thus $I_{\rm m}$ can be increased by raising either the dust mass or the temperature. However, increasing $T_{\rm d}$ requires larger effective optical depths $(\beta_{\rm UV} - \beta_{\rm int})$ (see also Sec. \ref{tdcosmic}), which are excluded in a relatively transparent single zone medium. It is possible to raise $M_{\rm d}$ while keeping $(\beta_{\rm UV} - \beta_{\rm UV,int})$ low, but this implies pushing the dust temperatures progressively closer to $T_{\rm CMB}$, thus preventing $F_{1900}$, and hence $I_{\rm m}$, to increase indefinitely. 
Interestingly, we find that the only galaxies in the ALPINE sample for which $I_{\rm m}>I_{\rm m}^*$ are the outliers in Fig. \ref{SFRcomp} (see also Tab. \ref{tab:RES_DET} for the $I_{\rm m}$ value of each source), with the largest value $I_{\rm m}=6752$ corresponding to the most peculiar galaxy DC 873756. 

These large $I_{\rm m}$ values\footnote{$I_{\rm m}>I_{\rm m}^*$ are also measured in few REBELS galaxies ($4$ out of the $14$ continuum detected sources)} can be achieved only if the FIR and UV emitting regions are spatially decoupled \citep{ferrara:2022}. In this scenario, the observed strong FIR continuum emission in these peculiar ALPINE galaxies comes from optically thick, star-forming clumps (likely, giant molecular complexes), whereas the small UV optical depth (i.e. blue $\beta_{\rm UV}$) traces the diffuse, interclump gas component in which young stars are embedded after they disperse their natal cloud \citep[see also][]{2017faisst}. This scenario, referred to as \quotes{spatial-segregation} of UV and IR emission has been proposed by other theoretical works focusing on sources at the EoR \citep{behrens18,Liang19,sommovigo20,pallottini:2022,dayal:2022}. 
Current ALMA observations do not to probe the ISM morphology down to such scales ($\ll 1\ \mathrm{kpc}$). Future ALMA observations at higher spatial resolution, combined with JWST images \citep[see e.g. $\propto 100\ \mathrm{pc}$ resolution images of $z=6-8$ galaxies by][]{2022arXiv220712657C} will help us confirm or discard the scenario proposed here. 

So far, a significant spatial offset between ALMA and HST data has been observed in some $z>5$ star-forming galaxies \citep{2012Hodge,carniani:2017oiii,laporte:2017apj,inami:2022,bowler2018obscured}. At lower redshift, spatial-segregation has been invoked to motivate the blue $\beta_{\rm UV}$ slopes found in some $z\sim 2$ dusty star forming galaxies (DSFGs, \citealt{2014casey}), which strongly deviate from the local IRX-$\beta$ relation (where $\mathrm{IRX}=L_{\rm FIR}/L_{\rm UV}$). Similarly, the IRX excess observed in some $z\sim 2$ starbursts \citep{2018Elbaz} and $z\sim 4.5$ sub-mm galaxies \citep{2018ApJ...856..121G} has been associated to spatially decoupled UV and IR emission in resolved ALMA and HST maps.


\section{Summary}\label{summary}
In this paper, we study the dust continuum emission properties of the Far Infrared (FIR) continuum detected $z=5$ galaxies from the ALMA Large Program ALPINE \citep{lefevre:2020,FaisstALPINE}. To derive their dust temperature $T_{\rm d}$ using the single available FIR continuum observation (at rest-frame $158\ \mathrm{\mu m}$), we apply the method presented in \cite{Sommovigo:2021}. This method relies on the combination of the dust continuum flux with the overlying \CII\ emission line luminosity; $L_{\rm CII}$ serves as a proxy for the dust mass $M_{\rm d}$, breaking the degeneracy between $M_{\rm d}$ and $T_{\rm d}$ in the FIR SED fitting procedure. 

Having constrained $T_{\rm d}$ and $M_{\rm d}$ for all the ALPINE galaxies, we can uniquely derive dust-temperature-dependent properties such as the IR luminosity $L_{\rm IR}$ and obscured SFR fraction $\mathrm{SFR_{\rm IR}}$.
Providing this insight for ALPINE galaxies is fundamental as they constitute the most abundant sample of dusty star-forming galaxies at $z>4$. Moreover, using the results from \cite{Sommovigo:2022}, we can compare them with the higher-$z$ REBELS galaxies \citep{bouwens:2021}, consistently investigating the cosmic evolution of dust properties in an unprecedentedly large number of sources from $z\sim 5$ to $z\sim 7$. 

We summarize below our findings:
\begin{itemize}
    \item 
    The dust temperatures for ALPINE galaxies vary within the range $25-60\ \mathrm{K}$. The average value $\left< T_{\rm d} \right> = 48 \pm 8\ \mathrm{K}$ is consistent with the result from stacked SED fitting by \cite{bethermin_alpine}. It also matches the predictions from the physical model for the cosmic $T_{\rm d}$ evolution by \cite{Sommovigo:2022}, which finds a mild increase in the dust temperature with redshift $T_{\rm d} \propto (1+z)^{0.4}$;
    
    \item Dust masses for ALPINE galaxies are in the $0.6-25.1\ \times 10^{7}\ \mathrm{M_{\odot}}$ range. Due to the $\sim 2$ times warmer $T_{\rm d}$ that we infer for ALPINE galaxies, $M_{\rm d}$ are $7$ times lower than previously reported by \cite{2021Pozzi}. Thus, the resulting dust yields $\left< M_{\rm d}/M_{\star} \right> = 2\times 10^{-3}$ are now consistent with theoretical dust production constraints at $z\sim 5$. In particular, we do not need to invoke extreme scenarios, e.g. saturated grain growth or no dust destruction, which might instead be needed at $z\sim 7$ for a few peculiar REBELS galaxies \citep[see also][]{dayal:2022}. 
    
    \item The linear $M_{\rm d}-M_{\star}$ relation predicted by theoretical models is consistent with our results, differently from what is found at $z\sim 7$ for REBELS galaxies where it appears flat. This might be evidence of a rapid evolution of the $M_{\rm d}-M_{\star}$ relation at $z> 5$. However, probing a wider stellar mass range at both redshifts is needed to completely exclude that this is due to an observational bias; 

    \item We find $8$ ALPINE sources with $L_{\rm IR}>10^{12}\ \mathrm{L_{\odot}}$, comparable to Ultra-Luminous InfraRed Galaxies (ULIRGs). Among these $8$ ULIRGs-like sources, there are $4$ extreme systems where $\mathrm{SFR_{\rm IR}}>100\ \mathrm{M_{\odot}/yr}$, exceeding by a factor $> 3$ the total SFR deduced from UV-to-optical SED fitting. These outliers are the only sources showing large molecular index values $I_{\rm m}=(F_{\rm 1900}/F_{\rm UV})/(\beta_{\rm UV}-\beta_{\rm int})>1120$, the critical value for a single phase ISM \citep{ferrara:2022}. We thus predict that these outliers are spatially-segregated systems, where FIR emission comes from clumpy giant molecular clouds whereas the UV arises from the diffuse, UV transparent ISM. 

\end{itemize}

High-resolution observations at sub-kpc scales for both the UV \citep[such as][]{2022arXiv220712657C} and the dust continuum (also including shorter-wavelengths ALMA bands 8 and 9) of the ALPINE spatially-segregated galaxies will help us clarify the morphology of their ISM. An immediate improvement will be provided by upcoming high-resolution ALMA band 6 observations within the CRISTAL large program (PI: Herrera-Camus) and by JWST, whose pointings include $2$ of these $4$ peculiar sources. 

\begin{table*}
\begin{center}
\caption{Measured properties of the $21$ \CII\ and continuum detected ALPINE galaxies together with  adopted relative errors ($[z,L_{\rm CII}, F_{\rm 1900}]$ from \citealt{bethermin_alpine}, $[M_{\star},M_{\rm UV}, \beta_{\rm UV}]$ from \citealt{FaisstALPINE}, and $[r_{\rm CII}, r_{\star}]$ from \citealt{Fujimoto_alpine}). The value of $r_{\rm CII}$ derived in this work is marked with a $^{*}$ (see Sec. \ref{method} for the details). We also show the molecular index value $I_{\rm m}$ (see eq. \ref{Im}) directly derived from the data.}
\begin{tabular}{lccccccccccc}
\hline\hline
\multicolumn{10}{c}{\code{\CII\ AND CONTINUUM DETECTED ALPINE GALAXIES: DATA}}\\

name & $z$ & $L_{\rm CII}$ & $F_{\rm 1900}$ & $\log M_{\star}$  & $r_{\rm CII}$ & $r_{\star}$ & $M_{\rm UV}$ & $\beta_{\rm UV}$ & $I_{\rm m}$\\
\hline
   &      &   [$10^8\ \mathrm{L_{\odot}}$]       & [$\mu$Jy]   & [$\mathrm{M_{\odot}}$] & [kpc] &  [kpc]  & [mag] & \\
\hline
CG 32  & $4.4105$ &$8.2 \pm 0.8$ &$230 \pm 65$ &$9.75^{+0.24}_{-0.29}$ &$1.94 \pm 0.3$ &$0.91 \pm 0.13$ & $-21.274$ & $-0.859157$ & $397$\\[1mm]
DC 396844& $4.5424$ &$11.5 \pm 1.0$ &$346 \pm 69$ &$9.86^{+0.14}_{-0.19}$ &$2.56 \pm 0.33$ &$0.58 \pm 0.2$ & $-21.665$ & $-1.38062$ & $658$\\[1mm]
DC 417567 &$5.6700$ &$3.1 \pm 0.5$ &$201 \pm 60$ &$9.81^{+0.18}_{-0.11}$ &$2.07 \pm 0.58$ &$0.65 \pm 0.20$ & $-22.919$ & $-1.86659$ & $320$\\[1mm]
DC 422677 &$4.4381$ &$4.2 \pm 0.7$ &$375 \pm 123$ &$9.85^{+0.14}_{-0.16}$ &$1.10 \pm 0.50$ &$0.58 \pm 0.14$ & $-21.634$ & $-1.2423$ & $624$\\[1mm]
DC 488399 & $5.6704$ &$10.8 \pm 0.5$ &$252 \pm 32$ &$10.20^{+0.13}_{-0.15}$ &$1.32 \pm 0.16$ &$0.47 \pm 0.32$ & $-22.058$ & $-1.88283$ & $913$\\[1mm]
DC 493583 & $4.5134$ &$4.3 \pm 0.6$ &$235 \pm 81$ &$9.61^{+0.15}_{-0.11}$ &$1.89 \pm 0.51$ &$0.64 \pm 0.17$ & $-21.765$ & $-2.01243$ & $1051$\\[1mm]
\textbf{DC 494057(HZ4)} & $5.5448$ &$7.2 \pm 0.5$ &$179 \pm 30$ &$10.15^{+0.13}_{-0.15}$ &$2.48 \pm 0.25$ &$0.88 \pm 0.16$ & $-22.373$ & $-1.87832$ & $466$ \\[1mm]
DC 539609 & $5.1818$ &$4.9 \pm 0.6$ &$187 \pm 54$ &$9.38^{+0.12}_{-0.12}$ &$1.65 \pm 0.43$ &$0.77 \pm 0.16$ & $-22.357$ & $-2.20637$ & $1179$\\[1mm]
DC 683613 & $5.5420$ &$7.8 \pm 0.7$ &$245 \pm 54$ &$10.17^{+0.14}_{-0.15}$ &$1.82 \pm 0.33$ &$0.57 \pm 0.24$ & $-21.428$ & $-1.3048$ & $729$ \\[1mm]
\textbf{DC 848185(HZ6)} & $5.2931$ &$16.0 \pm 0.95$ &$319 \pm 50$ &$10.37^{+0.08}_{-0.19}$ &$3.47 \pm 0.25$ &$0.9 \pm 0.3$ & $-22.54$ & $-1.14217$ & $277$\\[1mm]
DC 881725 & $4.5777$ &$6.87 \pm 0.63$ &$349 \pm 90$ &$9.96^{+0.16}_{-0.11}$ &$2.26 \pm 0.33$ &$0.67 \pm 0.21$ & $-21.553$ & $-1.20243$ & $634$\\[1mm]
VC 5100822662 & $4.5205$ &$7.86 \pm 0.65$ &$210 \pm 38$ &$10.17^{+0.13}_{-0.14}$ &$2.59 \pm 0.37$ &$1.32 \pm 0.33$ & $-21.891$ & $-1.31549$ & $303$\\[1mm]
VC 5100969402 & $4.5785$ &$5.23 \pm 0.55$ &$327 \pm 99$ &$10.00^{+0.12}_{-0.14}$ &$1.62 \pm 0.33$ &$0.59 \pm 0.15$ & $-21.53$ & $-1.94423$ & $1583$\\[1mm]
VC 5100994794 & $4.5802$ &$5.57 \pm 0.51$ &$117 \pm 36$ &$9.73^{+0.15}_{-0.13}$ &$1.86 \pm 0.32$ &$1.63 \pm 0.35$ & $-21.342$ & $-1.62524$ & $398$\\[1mm]
VC 5101209780 & $4.5701$ &$7.31 \pm 1.56$ &$311 \pm 112$ &$10.05^{+0.12}_{-0.12}$ &{$3.25 \pm 0.72$}$^{*}$ &$1.00 \pm 0.24$ & $-22.143$& $-1.91506$ & $803$\\[1mm]
VE 530029038 & $4.4298$ &$6.9 \pm 0.8$ &$125 \pm 58$ &$9.93^{+0.21}_{-0.12}$ &$2.76 \pm 0.65$ &$1.96 \pm 0.14$ & $-21.95$ & $-1.4904$ & $197$ \\[1mm]
DC 552206 & $5.5016$ &$15.2 \pm 1.1$ &$285 \pm 73$ &$10.58^{+0.14}_{-0.16}$ &$3.41 \pm 0.35$ &$0.96 \pm 0.64$ & $-22.642$ & $-0.977538$ & $211$\\[1mm]
DC 818760 & $4.5613$ &$43.0 \pm 1.7$ &$1077 \pm 130$ &$10.63^{+0.11}_{-0.10}$ &$2.59 \pm 0.16$ &$0.75 \pm 0.17$ & $-22.225$ & $-0.548038$ & $679$\\[1mm]
DC 873756 & $4.5457$ &$36.2 \pm 1.3$ &$1354 \pm 76$ &$10.25^{+0.08}_{-0.10}$ &$2.36 \pm 0.11$ &$1.08 \pm 0.43$ & $-20.869$ & $-1.59125$ & $6752$\\[1mm]
VC 5101218326 & $4.5739$ &$18.3 \pm 0.8$ &$462 \pm 79$ &$11.01^{+0.07}_{-0.05}$ &$2.37 \pm 0.15$ &$1.46 \pm 0.32$ & $-22.345$ & $-0.86014$ & $315$\\[1mm]
VC 5180966608& $4.5296$ &$13.7 \pm 1.1$ &$419 \pm 84$ &$10.82^{+0.12}_{-0.13}$ &$5.10 \pm 0.42$ &$0.59 \pm 0.24$ & $-21.743$ & $-0.826991$ & $479$\\[1mm]
\hline
\label{tab:DATA_DET}
\end{tabular}
\end{center}
\end{table*}

\begin{table*}
\begin{center}
\caption{Predicted properties of the $21$ \CII\ and continuum detected ALPINE galaxies, respectively: \CII-to-total gas conversion factor $\alpha_{\rm CII}$ (eq. \ref{alfahz}), dust temperature $T_{\rm d}$ and mass $M_{\rm d}$, IR luminosity $\log L_{\rm IR}$, SN dust yield $y_d$ and obscured SFR, $\mathrm{SFR}_{\rm IR} [M_\odot {\rm yr}^{-1}] = 10^{-10}\ \mathrm{L_{\rm IR}} [L_\odot]$ \citep{kennicutt1998}.}
\begin{tabular}{lccccccc}
\hline\hline
\multicolumn{7}{c}{\code{\CII\ AND CONTINUUM DETECTED ALPINE GALAXIES: RESULTS}}\\
name &$\alpha_{\rm CII}$ & $T_{\rm d}$ & $\log M_{\rm d}$ & $y_{\rm d}$ &  $\log L_{\rm IR}$& $\mathrm{SFR_{\rm IR}}$ \\
\hline
     & &  [K] &  [$M_{\odot}$] & [$M_{\odot}/\mathrm{SN}$] & [$L_{\odot}$] & [$M_{\odot}/yr$]\\
\hline
CG 32 &  $ 4 ^{+ 5 }_{- 2 }$ & $ 51 ^{+ 18 }_{- 15 }$ & $ 7.20 ^{+ 0.34 }_{- 0.25 }$ & $ 0.15 ^{+ 0.18 }_{- 0.07 }$ & $ 11.87 ^{+ 0.54 }_{- 0.56 }$ & $ 75 ^{+ 182 }_{- 54 }$\\[1mm]
DC 396844 & $ 12 ^{+ 11 }_{- 5 }$ & $ 37 ^{+ 12 }_{- 8 }$ & $ 7.75 ^{+ 0.31 }_{- 0.29 }$ & $ 0.41 ^{+ 0.43 }_{- 0.2 }$ & $ 11.6 ^{+ 0.45 }_{- 0.33 }$ & $ 40 ^{+ 73 }_{- 21 }$\\[1mm]
DC 417567 & $ 16 ^{+ 19 }_{- 6 }$ & $ 48 ^{+ 11 }_{- 13 }$ & $ 7.33 ^{+ 0.33 }_{- 0.24 }$ & $ 0.18 ^{+ 0.2 }_{- 0.08 }$ & $ 11.9 ^{+ 0.29 }_{- 0.46 }$ & $ 79 ^{+ 75 }_{- 51 }$\\[1mm]
DC 422677 & $ 9 ^{+ 11 }_{- 5 }$ & $ 54 ^{+ 11 }_{- 14 }$ & $ 7.24 ^{+ 0.32 }_{- 0.33 }$ & $ 0.13 ^{+ 0.14 }_{- 0.07 }$ & $ 12.11 ^{+ 0.16 }_{- 0.47 }$ & $ 128 ^{+ 58 }_{- 84 }$\\[1mm]
DC 488399 & $ 5 ^{+ 6 }_{- 2 }$ & $ 53 ^{+ 16 }_{- 15 }$ & $ 7.38 ^{+ 0.34 }_{- 0.22 }$ & $ 0.08 ^{+ 0.1 }_{- 0.03 }$ & $ 12.2 ^{+ 0.47 }_{- 0.5 }$ & $ 157 ^{+ 308 }_{- 108 }$\\[1mm]
DC 493583 & $ 10 ^{+ 11 }_{- 4 }$ & $ 48 ^{+ 15 }_{- 14 }$ & $ 7.25 ^{+ 0.33 }_{- 0.24 }$ & $ 0.23 ^{+ 0.26 }_{- 0.1 }$ & $ 11.79 ^{+ 0.45 }_{- 0.54 }$ & $ 61 ^{+ 111 }_{- 44 }$\\[1mm]
DC 494057 \textbf{(HZ4)}  & $ 8 ^{+ 9 }_{- 3 }$ & $ 47 ^{+ 16 }_{- 12 }$ & $ 7.37 ^{+ 0.35 }_{- 0.26 }$ & $ 0.09 ^{+ 0.11 }_{- 0.04 }$ & $ 11.83 ^{+ 0.5 }_{- 0.45 }$ & $ 68 ^{+ 145 }_{- 44 }$\\[1mm]
DC 539609 & $ 6 ^{+ 7 }_{- 2 }$ & $ 55 ^{+ 14 }_{- 15 }$ & $ 7.13 ^{+ 0.31 }_{- 0.22 }$ & $ 0.30 ^{+ 0.31 }_{- 0.12 }$ & $ 12.0 ^{+ 0.36 }_{- 0.53 }$ & $ 101 ^{+ 132 }_{- 71 }$\\[1mm]
DC 683613 & $ 8 ^{+ 10 }_{- 3 }$ & $ 49 ^{+ 15 }_{- 13 }$ & $ 7.43 ^{+ 0.35 }_{- 0.24 }$ & $ 0.10 ^{+ 0.12 }_{- 0.04 }$ & $ 12.01 ^{+ 0.48 }_{- 0.48 }$ & $ 102 ^{+ 206 }_{- 68 }$\\[1mm]
DC 848185 \textbf{(HZ6)}  & $ 11 ^{+ 13 }_{- 5 }$ & $ 39 ^{+ 13 }_{- 9 }$ & $ 7.86 ^{+ 0.36 }_{- 0.3 }$ & $ 0.16 ^{+ 0.21 }_{- 0.08 }$ & $ 11.86 ^{+ 0.48 }_{- 0.34 }$ & $ 73 ^{+ 145 }_{- 40 }$\\[1mm]
DC 881725 & $ 11 ^{+ 12 }_{- 4 }$ & $ 47 ^{+ 14 }_{- 13 }$ & $ 7.49 ^{+ 0.34 }_{- 0.24 }$ & $ 0.18 ^{+ 0.22 }_{- 0.08 }$ & $ 11.96 ^{+ 0.46 }_{- 0.49 }$ & $ 91 ^{+ 172 }_{- 62 }$\\[1mm]
VC 5100822662  & $ 5 ^{+ 5 }_{- 2 }$ & $ 50 ^{+ 18 }_{- 14 }$ & $ 7.2 ^{+ 0.35 }_{- 0.25 }$ & $ 0.06 ^{+ 0.07 }_{- 0.02 }$ & $ 11.86 ^{+ 0.54 }_{- 0.52 }$ & $ 72 ^{+ 178 }_{- 51 }$\\[1mm]
VC 5100969402 & $ 9 ^{+ 11 }_{- 4 }$ & $ 52 ^{+ 13 }_{- 14 }$ & $ 7.33 ^{+ 0.32 }_{- 0.24 }$ & $ 0.11 ^{+ 0.13 }_{- 0.05 }$ & $ 12.06 ^{+ 0.33 }_{- 0.53 }$ & $ 116 ^{+ 133 }_{- 81 }$\\[1mm]
VC 5100994794 & $ 2 ^{+ 3 }_{- 1 }$ & $ 60 ^{+ 19 }_{- 19 }$ & $ 6.77 ^{+ 0.34 }_{- 0.23 }$ & $ 0.06 ^{+ 0.07 }_{- 0.02 }$ & $ 11.87 ^{+ 0.48 }_{- 0.64 }$ & $ 75 ^{+ 153 }_{- 58 }$\\[1mm]
VC 5101209780& $ 11 ^{+ 13 }_{- 4 }$ & $ 42 ^{+ 16 }_{- 12 }$ & $ 7.53 ^{+ 0.35 }_{- 0.28 }$ & $ 0.16 ^{+ 0.2 }_{- 0.08 }$ & $ 11.72 ^{+ 0.55 }_{- 0.5 }$ & $ 52 ^{+ 135 }_{- 36 }$\\[1mm]
VE 530029038 & $ 3 ^{+ 4 }_{- 1 }$ & $ 48 ^{+ 21 }_{- 16 }$ & $ 6.97 ^{+ 0.35 }_{- 0.27 }$ & $ 0.06 ^{+ 0.07 }_{- 0.03 }$ & $ 11.5 ^{+ 0.66 }_{- 0.69 }$ & $ 31 ^{+ 113 }_{- 25 }$\\[1mm]
DC 552206 & $ 10 ^{+ 12 }_{- 4 }$ & $ 37 ^{+ 13 }_{- 9 }$ & $ 7.79 ^{+ 0.37 }_{- 0.3 }$ & $ 0.09 ^{+ 0.12 }_{- 0.04 }$ & $ 11.66 ^{+ 0.48 }_{- 0.34 }$ & $ 45 ^{+ 93 }_{- 25 }$\\[1mm]
DC 818760 & $ 6 ^{+ 7 }_{- 2 }$ & $ 45 ^{+ 16 }_{- 12 }$ & $ 8.04 ^{+ 0.36 }_{- 0.28 }$ & $ 0.14 ^{+ 0.17 }_{- 0.06 }$ & $ 12.38 ^{+ 0.53 }_{- 0.45 }$ & $ 240 ^{+ 575 }_{- 154 }$\\[1mm]
DC 873756 & $ 5 ^{+ 5 }_{- 1 }$ & $ 59 ^{+ 14 }_{- 16 }$ & $ 7.88 ^{+ 0.32 }_{- 0.17 }$ & $ 0.22 ^{+ 0.24 }_{- 0.07 }$ & $ 12.94 ^{+ 0.38 }_{- 0.55 }$ & $ 878 ^{+ 1233 }_{- 628 }$\\[1mm]
VC 5101218326 & $ 3 ^{+ 3 }_{- 1 }$ & $ 59 ^{+ 17 }_{- 18 }$ & $ 7.40 ^{+ 0.34 }_{- 0.2 }$ & $ 0.01 ^{+ 0.02 }_{- 0.0 }$ & $ 12.48 ^{+ 0.46 }_{- 0.59 }$ &$ 303 ^{+ 569 }_{- 225 }$\\[1mm]
VC 5180966608 & $ 47 ^{+ 54 }_{- 21 }$ & $ 25 ^{+ 6 }_{- 4 }$ & $ 8.41 ^{+ 0.35 }_{- 0.31 }$ & $ 0.20 ^{+ 0.26 }_{- 0.1 }$ & $ 11.23 ^{+ 0.22 }_{- 0.12 }$ & $ 17 ^{+ 11 }_{- 4 }$\\[1mm]
\hline
\label{tab:RES_DET}
\end{tabular}
\end{center}
\end{table*}

\section*{Acknowledgements}
AF, AP, LS, SC acknowledge support from the ERC Advanced Grant INTERSTELLAR H2020/740120 (PI: Ferrara). Any dissemination of results must indicate that it reflects only the author’s view and that the Commission is not responsible for any use that may be made of the information it contains. Partial support from the Carl Friedrich von Siemens-Forschungspreis der Alexander von Humboldt-Stiftung Research Award is kindly acknowledged (AF). PD acknowledges support from the ERC starting grant DELPHI StG-717001, from the NWO grant ODIN 016.VIDI.189.162 and the European Commission's and University of Groningen's CO-FUND Rosalind Franklin program.

\section*{Data Availability}
Data generated in this research will be shared on reasonable request to the corresponding author.

\bibliographystyle{stile/mnras}
\bibliography{bibliografia/bibliogr,bibliografia/codes}

\bsp 

\label{lastpage}
\end{document}